\documentclass{article}

\usepackage{arxiv}

\usepackage[utf8]{inputenc} 
\usepackage[T1]{fontenc}    
\usepackage{hyperref}       
\usepackage{url}            
\usepackage{booktabs}       
\usepackage{amsfonts}       
\usepackage{nicefrac}       
\usepackage{microtype}      
\usepackage{lipsum}		
\usepackage{graphicx}
\usepackage{natbib}
\usepackage{doi}
\usepackage{amsmath,amsfonts,amssymb}
\usepackage{comment}

\DeclareUnicodeCharacter{03B2}{\ensuremath{\beta}}

\title{ADDA: a Modular Framework for Representing, Simulating and Assimilating Dynamics with End-to-end Differentiability}

\date{} 					

\author{ 
Anthony Frion\thanks{Corresponding author}\\
  \texttt{anthony.frion@hereon.de}\\ 
  \And
 Vien Minh Nguyen-Thanh\\
  \texttt{thanh.nguyen@hereon.de} \\
  \And
 Ali Can Bekar\\
  \texttt{ali.bekar@hereon.de} \\
  \And
 Pauleo R. Nimtz\\
  \texttt{pauleo.nimtz@hereon.de} \\
  \And
 Vadim Zinchenko\\
  \texttt{vadim.zinchenko@hereon.de} \\
   \And
 David S. Greenberg\\
  \texttt{david.greenberg@hereon.de} \\
    \And 
    \\
    Model-Driven Machine Learning\\
  Institute of Coastal Systems - Analysis and Modeling\\
  Helmholtz-Zentrum Hereon\\
  Germany\\
}
\renewcommand{\shorttitle}{Combined Optimization of Dynamics and Assimilation}

\begin{document}
\maketitle
\begin{abstract}
Data assimilation (DA) is an essential tool for prediction and understanding in the geosciences. DA combines simulation programs representing scientific knowledge with observations that constrain system dynamics, resulting in analyses and forecasts that incorporate both knowledge and data. DA tasks can be addressed with a diverse toolset, including variational, ensemble and learning-based methods. In particular, many recent works have proposed using automatic differentiation tools for variational, learning-based or hybrid methods. However, comprehensive comparisons across algorithms and dynamical systems remain challenging, due to the incompatibility of simulation and assimilation codes, inflexible handling of spatial and temporal discretizations, specialization of DA methods to specific simulations, and limited support for automatic differentiation and parallel computation in simulations. To address this challenge, we introduce Automatic Differentiation for Data Assimilation (ADDA), a software framework for defining and working with system states, simulations, observation schemes and DA methods. ADDA provides a powerful and flexible set of base classes for representing dynamical systems and observation operators, with support for collocated and staggered grids, unstructured meshes, Lagrangian state variables and irregular or continuous-time observations. ADDA implements the Kalman filter and smoother, ensemble Kalman filter and smoother, strong-constraint and weak-constraint 4D-Var with a single or a sliding window. It can be easily extended or modified to support new methods. Parallel processing and differentiability are first-class features, with support for batch axes and automatic differentiation throughout. ADDA is implemented in PyTorch library, but supports DA for JAX-based computation of dynamics and their gradients. To demonstrate its features and facilitate development and evaluation of DA methods, we further provide differentiable, ADDA-compatible implementations of 10 dynamical systems of various dimensionalities and scales, from which we design multiple illustrative DA examples. All of our code is publicly available at \url{https://github.com/m-dml/ADDA}.
\end{abstract}


\section{Introduction}
Data assimilation (DA)~\citep{carrassi_data_2018}, a critically important endeavor in the geosciences, combines simulation and observation to obtain system state sequences consistent with both. It can be used to estimate system states more accurately than simulation or observation alone~\citep{carrassi_data_2018,hersbach_era5_2020}, to tune simulation parameters~\citep{aksoy_ensemble-based_2006, bocquet_joint_2013} and to identify key situations or phenomena for which simulation and observations conflict~\citep{carrio_high-resolution_2025}. DA is the focus of much theoretical and empirical research, and its operational practice demands significant financial, computational and energy resources~\citep{balsamo_recent_2023, drillet_core_2024}.

DA unites a diverse suite of methods with an equally varied array of applications. Major classes of well-established ``classical'' methods include variational methods such as strong-~\citep{rabier_ecmwf_2000} and weak-constraint 4D-Var~\citep{zupanski_general_1997}, 
gradient-free ensemble methods~\citep{evensen_sequential_1994, evensen_ensemble_2000} and hybrid techniques combining gradients with ensembles~\citep{liu_ensemble-based_2008, sakov_iterative_2012}. Kalman filtering provides precise Bayesian inference for linear dynamics with Gaussian uncertainty~\citep{kalman_new_1960}, while particle filters can assimilate nonlinear dynamics or observations but suffer from a curse of dimensionality that limits their application in the geosciences~\citep{wills_sequential_2023}. Within these classes of DA methods, individual methods exhibit subtle but consequential differences, and implementations of a single method can vary widely in usability, flexibility, numerical stability and computational efficiency.

Advances in data-driven DA present an important, new and highly diverse class of methods~\citep{cheng_machine_2023}. Fully supervised approaches avoid simulation during training, but generally require simulation or another DA method to produce training data~\citep{hersbach_era5_2020}, in the form of full system states. Directly training on sparse and/or noisy observations usually assumes access to a differentiable simulation~\citep{zinchenko_combined_2024,wang_four-dimensional_2024} or emulator~\citep{xiao_fengwu-4dvar_2024,li_fuxi-en4dvar_2024}. 4DVarNets~\citep{fablet_learning_2021} learn a 4D-Var-like loss and pass its gradients to a recurrent DA network, supporting both supervised~\citep{beauchamp_4dvarnet-ssh_2023,febvre_training_2024,fablet_inversion_2024} and unsupervised training~\citep{dorffer_observation-only_2025}. Other approaches train a fast neural emulator on simulations that are too slow or lack gradient routines, enabling classical variational DA~\citep{nonnenmacher_deep_2021, hatfield_building_2021, maulik_efficient_2022, seabra_ai_2024, xiao_fengwu-4dvar_2024, li_fuxi-en4dvar_2024,durand_four-dimensional_2025}. In some cases, emulators trained on coarse-resolution 
simulations exhibit better consistency with high-resolution simulations than the coarse simulations they were trained on~\citep{koehler_neural_2026}. However, emulation is unnecessary when simulations are implemented in deep learning frameworks with automatic differentiation (autodiff), since gradient backpropagation is closely related to variational DA~\citep{cheng_machine_2023}. Latent data assimilation (LDA) reduces computation and memory costs by operating in latent spaces to reduce the dimension~\citep{amendola_data_2020, peyron_latent_2021, cheng_multi-domain_2024, melinc_3d-var_2024, cheng_torchda_2025}, to leverage simplified (e.g. linear) dynamics~\citep{frion_neural_2024,singh_koda_2024, frion_augmented_2025, shoji_poster_2025,tong_latent_2026} or to work with complex non-Gaussian prior distributions~\citep{qu_deep_2024, xiao_vae-var_2024, xiao_lora-envar_2025, pasmans_ensemble_2026, fan_physically_2026}.
Diffusion models have been used for probabilistic DA, both by directly mapping observations onto system state posteriors~\citep{huang_diffda_2024} and by training a prior on system state sequences~\citep{rozet_score-based_2023} before conditioning with Tweedie's formula~\citep{efron_tweedies_2011}. Diffusion models and related approaches can incorporate latent representations~\citep{qu_deep_2024}, exploit novel training strategies~\citep{shysheya_conditional_2024, chen_flowdas_2025, sun_align-da_2025}, learn from observations~\citep{rozet_learning_2024} and scale to global atmospheric states~\citep{andry_appa_2025}. Overall, data-driven DA methods are nearly always faster than classical DA, and can be used to initialize iterative classical methods~\citep{frerix_variational_2021, frion_uncertainty-aware_2026}. 

The complex landscape of DA applications is at least as varied as that of its methods. These include the familiar cases of atmospheric~\citep{navon_data_2009, gustafsson_survey_2018, hersbach_era5_2020, xu_fuxi-da_2025} and ocean dynamics~\citep{stammer_ocean_2016,drillet_core_2024,grande_machine_2026}, as well as hydrology~\citep{sun_review_2016}, ecosystems~\citep{niu_role_2014}, biogeochemistry~\citep{dowd_statistical_2014, rayner_fundamentals_2019}, agriculture~\citep{jin_review_2018,pandya_review_2022}, 
pose estimation~\citep{aharon_uncertainty-aware_2025}, economics~\citep{nadler_data_2019}, water quality modeling~\citep{cho_data_2020}, forest inventory~\citep{xu_harnessing_2023}, land surface modeling~\citep{raoult_parameter_2025} and agent-based modeling~\citep{ghorbani_data_2023}. The simulation models underlying these diverse tasks rely on diverse programming languages, compilers, computing hardware, discretization schemes and data formats. Software packages for some simulations include their own assimilation routines~\citep{brajard_combining_2021,xiao_vae-var_2024,xu_fuxi-da_2025}, but these are often highly specialized and cannot be applied to other cases.

This complexity and diversity limit compatibility. Researchers developing a new DA method must contend with the difficulty of integrating it with diverse simulation codebases, system representations and observation types. Likewise, comparing assimilation methods on a new (or newly modified) simulation requires adapting the simulation to provide inputs with the format and structure required by each algorithm, and translating observation operators to an accepted format. When simulations support parallel or gradient computation, extending these capabilities to DA methods is nontrivial. The laborious nature of matching DA methods to simulations in each case costs precious research time, limits the feasibility of careful evaluations and increases the risk of implementation errors. Thus, at present it is a significant challenge to evaluate a new DA method, compare DA methods across assimilation tasks, or select the optimal method for each task.

Here we introduce ADDA, a modular framework for representing and simulating dynamical systems, for implementing observation operators, and for building, applying and evaluating data assimilation methods. We built ADDA to satisfy the following set of core requirements:
\begin{itemize}
    \item End-to-end autodiff support for variational DA methods.
    \item Support for diverse system state structures, including regular and unstructured grids and meshes, variable staggering, and mixed dimensionality (e.g. surface vs. atmospheric fields for Earth models).
    \item Observation operators describing the dependence of observations on system states, supporting spatiotemporally structured and unstructured observations, masking and noise.
    \item Support for flexible classes of dynamics, including time-varying boundary conditions and forcing data and numerical simulators as well as neural emulators.
    \item Modularity allowing simulation models, observation operators and DA methods to be individually defined, modified and extended before being combined to carry out DA tasks.
    \item Parallelized batched computation on CPUs and GPUs.
    \item Concrete, easily-modified examples demonstrating major functionalities.
\end{itemize}
We implemented ADDA as a set of algorithms, functions and classes in Python and PyTorch. ADDA satisfies our stated core requirements and provides a clearly defined interface through which simulations and DA algorithms can interoperate. 


ADDA is inspired by existing efforts to unify DA models, observation operators and simulations in a common framework. DART~\citep{anderson_data_2009} includes many types of dynamics and observation types, but focuses on ensemble-based DA and lacks variational methods. Its Fortran implementation aids efficiency but increases the complexity of its workflows and of interfacing with new simulations. PDAF~\citep{nerger_pdaf_2023} applies ensemble Kalman methods to large scale dynamical systems with emphasis on parallelization, but does not support 4D-Var.
NEDAS~\citep{ying_nansencenternedas_2025} is an ensemble DA system in Python with CPU parallelization, designed to efficiently scale to large geophysical applications. However, it lacks automatic differentiation, and its required workflows are complex and incompletely documented. PyDA~\citep{ahmed_pyda_2020} provides a simple data assimilation tutorial in Python, but also lacks support for automatic differentiation, so its variational DA requires explicit derivation and implementation of cost function gradients, which can become cumbersome for more complex systems.
DAPPER~\citep{raanes_dapper_2024} is a Python package implementing a variety of variational, ensemble and other methods, with many examples reproducing published studies. It presents a major achievement in unifying DA methods and applications, but with significant limitations: it is not designed for large systems (>60000 DOF), uses inefficient user-specified forward adjoints instead of autodiff backpropagation, provides only limited CPU parallelization and imposes restrictions on the spatial discretization, time stepping and observation scheme.  TorchDA~\citep{cheng_torchda_2025} supports autodiff in data assimilation, but only for a limited set of methods and examples focused on DA with neural emulators of dynamical systems. Each of these frameworks serves an important community need, but none fulfill the above aims. 
Generally speaking, one can note that variational methods have historically been considered to be complex to implement due to the necessity of building an adjoint model for the dynamical system of interest, leading to a preference for ensemble Kalman methods in general-purpose data assimilation software. The recent development of automatic differentiation tools has now considerably simplified the usage of 4D-Var, but this change is not yet reflected in major DA packages. Our work aims to fill this gap. As multiple modern frameworks exist for automatic differentiation, ADDA is implemented in PyTorch, the most popular for deep learning. To support a wide and flexible usage, we provide support for solving DA on systems implemented in JAX, which is another major automatic differentiation framework. An associated experiment is presented in Sect.~\ref{sec:ksjax4dvar}.

We emphasize that ADDA is not a benchmark or benchmarking tool, and this manuscript does not describe a benchmarking study. Arriving at a fair and comprehensive benchmark will require a concerted community effort to ensure that all relevant DA methods are adequately implemented and properly tuned, that a fair and representative range of DA tasks has been selected, and that reasonable metrics have been used to measure accuracy and efficiency. Rather, we designed and developed ADDA to provide a necessary foundation for implementing DA methods and interfacing them with simulations, keeping in mind the goal of enabling future benchmarking efforts. Our set of examples in Sect.~\ref{sec:examples} has been designed to showcase the versatility of ADDA.

We describe and demonstrate ADDA as follows. We first frame the DA task and establish notation in Sect.~\ref{sec:problem_statement}. In Sect.~\ref{sec:methods}, we describe essential DA concepts and components and their realizations in ADDA, and the implementation of DA algorithms. Section~\ref{sec:examples} then illustrates ADDA's capabilities on DA examples ranging across dynamical systems, observation operators and algorithms. Sections~\ref{sec:discussion} and~\ref{sec:conclusion} conclude with a summary of ADDA's capabilities and the outlook for future developments and applications.

\section{Data Assimilation: Methods and Formulations}
\label{sec:problem_statement}
In this section we formalize the DA task. We begin by framing the ``standard'' setup of a nonlinear dynamical system with per-time-point observations as well as Gaussian process and observation noise (Sect.~\ref{sec:standardda}). We then derive and describe the Bayesian posterior on states given observations (Sect.~\ref{sec:bayes}), and how various DA approaches relate to it (Sect.~\ref{sec:4dvar} and~\ref{sec:sequential}). Beyond the standard setup, we consider a more general class of scenarios supported by ADDA in Sect.~\ref{sec:da_extensions}.
\subsection{DA task formulation: Gaussian process and observation noise} \label{sec:standardda}
We consider a state-space model evolving over discrete time steps~$t$ with Gaussian process and observation noise terms. Each observation~$\mathbf{y}_t$ depends on the simultaneous state~$\mathbf{x}_t$, according to the following equations:
\begin{align}
    \label{eq:state_equation}
    \mathbf{x}_{t+1} &= \mathcal{M}(\mathbf{x}_t) + \boldsymbol{\epsilon}_t, \quad \boldsymbol{\epsilon}_t \sim \mathcal{N}(0, \boldsymbol{\Sigma_\epsilon})\\
    \label{eq:observation_equation}
    \mathbf{y}_t &= \mathcal{H}_t(\mathbf{x}_t) + \boldsymbol{\eta}_t, \quad \boldsymbol{\eta}_t \sim \mathcal{N}(0, \boldsymbol{\Sigma}_{\boldsymbol{\eta}_t}) \\
    \mathbf{x}_0 &\sim \mathcal{N}(\boldsymbol{\mathbf{x}}^B, \boldsymbol{\Sigma}_B)
    \label{eq:Gaussianbackgroundprior}
\end{align}


Equations~\eqref{eq:state_equation} and~\eqref{eq:observation_equation} are respectively referred to as the state and observation equations. The state $\mathbf{x}_t \in \mathbb{R}^s$ evolves through the discrete dynamical model $\mathcal{M}: \mathbb{R}^s \to \mathbb{R}^s$, with zero-mean Gaussian process noise $\boldsymbol{\epsilon}_t$. Observations $\mathbf{y}_t \in \mathbb{R}^{o_t}$ depend on $\mathbf{x}_t$ through the observation operator $\mathcal{H}_t: \mathbb{R}^s \to \mathbb{R}^{o_t}$. Observation noise $\boldsymbol{\eta}_t \in \mathbb{R}^{o_t}$ follows a Gaussian distribution with mean 0 and covariance $\boldsymbol{\Sigma}_{\boldsymbol{\eta}_t}$. A simple but important case is the linear observation $\mathcal{H}_t(\mathbf{x}_t) = \mathbf{H}_t\mathbf{x}_t$, where~$\mathbf{H}_t$ is a ${o_t \times s}$ matrix. Clearly, $\mathcal H_t$ is not invertible when $o_t < s$.

In DA, our task is to estimate the system state trajectory~$\mathbf{x}_{0:T}$ from an observation sequence~$\mathbf{y}_{0:T}$. In addition to the observation sequence, we assume full knowledge of the dynamics $\mathcal M$, process noise~$\boldsymbol{\Sigma}_{\boldsymbol{\epsilon}}$, dependence of expected observations on states~$\mathcal H_t$, observation noise~$\boldsymbol{\Sigma}_{\boldsymbol{\eta}_t}$ and initial state prior~$p(\mathbf{x}_0)$.


\subsection{Bayesian derivation of the posterior distribution of the state}
\label{sec:bayes}
Given a sequence of observations, the full probabilistic solution to the DA problem describes the full range of possible system state sequences, and how likely they are given the available data. In practice, many effective DA methods approximate the full posterior distribution with a point estimate, a marginal distribution for each time point or an ensemble of finite size, or by taking into account only a subset of observations. We describe these in Sect.~\ref{sec:4dvar} and~\ref{sec:sequential}. The posterior distribution over states given observations derives from the prior, state and observation equations according to Bayes' theorem:
\begin{equation} \label{eq:bayes}
    p(\mathbf{x}_{0:T}|\mathbf{y}_{0:T}) = \frac{p(\mathbf{x}_{0:T})p(\mathbf{y}_{0:T}|\mathbf{x}_{0:T})}{p(\mathbf{y}_{0:T})}.
\end{equation}
Equation~\eqref{eq:state_equation} is an example of a state space model, which has the Markovian property that each state's distribution, when conditioned on all earlier states, depends only on the immediately preceding state, so that the prior on state sequences factorizes: 
\begin{equation}
\label{eq:prior_factorisation}
    p(\mathbf{x}_{0:T}) = p(\mathbf{x}_0)\prod_{t=1}^T p(\mathbf{x}_t|\mathbf{x}_{t-1}).
\end{equation}
Similarly, assuming that observations are independent in time and that each of them depends only on a single, simultaneously occurring state yields a factorized likelihood:
\begin{equation}
\label{eq:obs_error_factorisation}
    p(\mathbf{y}_{0:T}|\mathbf{x}_{0:T}) = \prod_{t=0}^T p(\mathbf{y}_t|\mathbf{x}_t).
\end{equation}
Combining these assumptions, and dropping the marginal likelihood $p(\mathbf{y}_{0:T})$ which is constant with respect to the unknown states $\mathbf{x}_{0:T}$, we arrive at the factorized posterior:
\begin{equation}
\label{eq:factorizedposterior}
 p(\mathbf{x}_{0:T}|\mathbf{y}_{0:T}) \propto
p(\mathbf{x}_0)\prod_{t=1}^T p(\mathbf{x}_t|\mathbf{x}_{t-1}) \prod_{t=0}^T p(\mathbf{y}_t|\mathbf{x}_t).
\end{equation}
Plugging in the Gaussian densities from Eq.~\eqref{eq:state_equation}-\eqref{eq:Gaussianbackgroundprior}, we arrive at:
\begin{equation}
\label{eq:standardposterior}
\log \, p(\mathbf{x}_{0:T}|\mathbf{y}_{0:T}) = -\frac{1}{2} \left(\left\lVert\mathbf{x}_0 - \mathbf{x}^B\right\rVert^2_{\boldsymbol{\Sigma}_B^{-1}} + 
\sum_{t=1}^T \left\lVert\mathbf{x}_t - \mathcal{M}(\mathbf{x}_{t-1})\right\rVert^2_{\boldsymbol\Sigma_{\boldsymbol\epsilon}^{-1}}+ 
\sum_{t=0}^T\left\lVert\mathbf{y}_t - \mathcal{H}_t(\mathbf{x}_t)\right\rVert^2_{\boldsymbol\Sigma_{\boldsymbol{\eta}_t}^{-1}}
\right)
+ C
\end{equation}
where $||\mathbf{x}||^2_{\boldsymbol{\Sigma}^{-1}} = ||\mathbf{x}^\intercal \mathbf{\Sigma}^{-1}\mathbf{x}||^2$ is a weighted inner product and~$C$ does not depend on~$\mathbf{x}_{0:T}$. Note that if the prior on the initial state~$\mathbf{x}_0$ is taken to be uniform instead of Gaussian, the first term of Eq.~\eqref{eq:standardposterior} vanishes.

For the case of nonlinear dynamics~$\mathcal M$, the posterior distribution is intractable and non-Gaussian, and calculating or sampling from this density is challenging. This motivates specialized DA algorithms that employ various approximations while exploiting the temporal structure of the assimilation task.

\subsection{The 4D-Var cost and its minimization}
\label{sec:4dvar}
4D-Var, the most common variational data assimilation method, is a \textit{maximum a posteriori} (MAP) technique that seeks the mode of the posterior distribution from Eq.~\eqref{eq:standardposterior}, by minimizing the generic variational cost~$J$:
\begin{equation}
\label{eq:wc4dvar}
    \underset{\mathbf{x}_{0:T} \in \mathbb{R}^{s \times (T+1)}}{\arg \min} J(\mathbf{x}_{0:T}) =  
    \underset{\mathbf{x}_{0:T} }{\arg \min} \quad 
    \frac{1}{2} \left(\left\lVert\mathbf{x}_0 - \mathbf{x}^B\right\rVert^2_{\boldsymbol{\Sigma}_B^{-1}} + 
    \sum_{t=1}^T \left\lVert\mathbf{x}_t - \mathcal{M}(\mathbf{x}_{t-1})\right\rVert^2_{\boldsymbol\Sigma_\epsilon^{-1}} + 
    \sum_{t=0}^T\left\lVert\mathbf{y}_t - \mathcal{H}_t(\mathbf{x}_t)\right\rVert^2_{\boldsymbol\Sigma_{\eta_t}^{-1}}\right). 
\end{equation}

The minimization problem of Eq.~\eqref{eq:wc4dvar} is classically referred to as \textit{weak-constraint} 4D-Var: see e.g.~\cite{tremolet_model-error_2007, bannister_review_2017, carrassi_data_2018, evensen_weak_2022} for similar expressions. It should however be noted that multiple alternative expressions exist: see e.g.~\cite{tremolet_model-error_2007,laloyaux_exploring_2020} for some examples. The first term in Eq.~\eqref{eq:wc4dvar} measures deviation of the initial state from prior expectations, the second measures deviation of the state's evolution from the expected dynamics (hence the name ``weak-constraint''), and the third measures deviation from observations. While computing the second term in Eq.~\eqref{eq:wc4dvar} requires~$T$ time steps of simulated dynamics~$\mathcal M$ to evaluate~$J$, it does not require autoregressive iteration of~$\mathcal M$, allowing straightforward parallel computation over the time axis.

The \textit{strong-constraint} formulation of 4D-Var (sometimes referred to simply as 4D-Var) instead requires that $\mathbf{x}_{t} = \mathcal M(\mathbf{x}_{t-1})$ precisely \citep{carrassi_data_2018}. Thus, the model error $\boldsymbol{\epsilon}_t=0$ in Eq.~\eqref{eq:state_equation}, and the state evolves deterministically from its initialization~$\mathbf{x}_0$. In this case, the variational cost in Eq.~\eqref{eq:wc4dvar} is redefined for optimization only over~$\mathbf{x}_0$:
\begin{equation}
\label{eq:sc4dvar}
    \underset{\mathbf{x}_{0} \in \mathbb{R}^{s}}{\arg \min} \quad J(\mathbf{x}_{0}) = 
    \underset{\mathbf{x}_{0} \in \mathbb{R}^s }{\arg \min}\quad    
    \frac{1}{2} \left(\left\lVert\mathbf{x}_0 - \mathbf{x}^B\right\rVert^2_{\boldsymbol{\Sigma}_B^{-1}} + 
    \sum_{t=0}^T\left\lVert\mathbf{y}_t - \mathcal{H}_t(\mathcal{M}^{(t)}(\mathbf{x}_0))\right\rVert^2_{\boldsymbol{\Sigma}_{\boldsymbol\eta_t}^{-1}}\right),
\end{equation}
where~$\mathcal{M}^{(t)}$ means~$t$ compositions of~$\mathcal{M}$. The effectiveness of weak- vs. strong-constraint 4D-Var can depend on how accurately we know~$\mathcal M$ and~$\boldsymbol{\Sigma}_{\boldsymbol{\eta}_t}$, the sensitivity of the dynamics to initial perturbations and the nature of available observations.

To minimize~$J$, weak- and strong-constraint 4D-Var methods rely on some form of iterative gradient-based optimization. This generally requires a so-called ``adjoint'' model of the dynamics, $\mathbf{M}=\frac{d\mathcal{M}}{d\mathbf{x}}$. As detailed in e.g.~\cite{bannister_elementary_2001}, the gradient of Eq.~\eqref{eq:sc4dvar} can then be written as
\begin{equation}
    \nabla_{\mathbf{x}_0}J = \mathbf{\Sigma}_B^{-1}(\mathbf{x}_0 - \mathbf{x}^B) - \sum_{t=0}^T (\mathbf{M}^\intercal)^{t}\mathbf{H}_t^\intercal\boldsymbol{\Sigma_\eta}^{-1}(\mathbf{y}_t - \mathcal{H}_t(\mathbf{x}_t)),
\end{equation}
where~$\mathbf{H}_t$ is a linearization of the observation operator~$\mathcal{H}_t$ at time~$t$. Deriving, implementing and maintaining the adjoint~$\mathbf{M}$ of a dynamical system~$\mathcal{M}$ using the chain rule is widely considered to be a cumbersome operation, making it the main disadvantage of 4D-Var with regards to ensemble Kalman methods. However, one major contribution of our proposed ADDA package is the substitution of the manual derivation of the adjoint~$\mathbf{M}$ by an automatic computation of this adjoint using automatic differentiation. Thus, given that the system~$\mathcal{M}$ of interest has a differentiable implementation in PyTorch or JAX, the usage of various 4D-Var variants becomes straightforward.
Although re-implementing classical models to make them auto-differentiable can also be a cumbersome process, it has repeatedly proven to be a successful approach~\citep{gelbrecht_differentiable_2023, zhou_proof--concept_2024, solvik_4d-var_2025, meunier_towards_2025}. Alternatively, a non-differentiable implementation of the dynamics~$\mathcal{M}$ can be replaced by a neural emulator~\citep{nonnenmacher_deep_2021,hatfield_building_2021,penny_integrating_2022, maulik_efficient_2022,durand_four-dimensional_2025}, which is then differentiable by design. 

\subsection{Sequential methods}
\label{sec:sequential}
Besides variational methods, the other major class of data assimilation techniques is sequential methods, which assimilate each observation individually while moving forward along the time axis. The most prominent sequential methods include Kalman filters~\citep{kalman_new_1960} and smoothers~\citep{rauch_maximum_1965} as well as ensemble Kalman filters~\citep{evensen_sequential_1994} and smoothers~\citep{evensen_ensemble_2000}. An important advantage of these methods compared to variational approaches is that they do not require an adjoint model of the state dynamics~$\mathcal{M}$. Unlike variational methods that optimize to find the posterior mode, at each time step a sequential method directly computes or approximates the \textit{filtering distribution} $p(\mathbf{x}_t|\mathbf{y}_{0:t})$, the posterior over states given past and present (but not future) observations. Filtering methods alternate between forecasting steps that simulate a single time step through~$\mathcal M$ to advance $p(\mathbf{x}_{t-1}|\mathbf{y}_{0:t-1}) \rightarrow p(\mathbf{x}_{t}|\mathbf{y}_{0:t-1})$, and analysis steps that condition on a new observation~$\mathbf{y}_t$ to obtain $p(\mathbf{x}_{t}|\mathbf{y}_{0:t})$. After filtering, a subsequent smoothing step can then incorporate future observations to compute the \textit{smoothing distribution} $p(\mathbf{x}_t|\mathbf{y}_{0:T})$. A \textit{fixed-lag smoother} incorporates observations from some fixed number~$d$ of future time steps to obtain $p(\mathbf{x}_t|\mathbf{y}_{0:t+d})$.

Filtering and smoothing distributions can be computed precisely only in a narrow range of circumstances. Kalman filters and smoothers do this for the linear-Gaussian case, where~$\mathcal M$ in Eq.~\eqref{eq:state_equation} and~$\mathcal H_t$ in Eq.~\eqref{eq:observation_equation} are linear operators. In this case the filtering, smoothing and fixed-lag smoothing distributions are Gaussian as well, and their means and covariances can be computed in closed form through iterative updates along the time axis.

In contrast, ensemble Kalman filters and smoothers represent probability distributions through discrete ensembles sampled from these distributions. This allows for accurate assimilation with nonlinear dynamics and observation operators in many practically important scenarios. In the forecast step, each ensemble member is advanced independently through the nonlinear dynamics~$\mathcal M$. In the analysis step, the empirical covariance of state variables across ensemble members is used to condition on observations similarly to Kalman filtering of linear-Gaussian systems. The resulting posterior distribution is realized by nudging the individual ensemble members towards observations, according to a deterministic or probabilistic update rule. Critically, the full state covariance matrix need not be explicitly calculated, enabling a significant reduction in the computational and memory costs for ensemble size~$<s$. However, approximating the state covariance from a small ensemble can produce spurious correlations, possibly causing the well-known problem of ``filter divergence''~\citep{houtekamer_data_1998}. This problem has been addressed through various techniques such as inflation~\citep{anderson_monte_1999} and localization~\citep{greybush_balance_2011}. 

Many ensemble filter and smoother variants exist. The ensemble adjustment Kalman filter~\citep{anderson_ensemble_2001} is a popular deterministic variant, belonging to the family of ensemble square root filters~\citep{tippett_ensemble_2003}. The iterative ensemble Kalman filter (IEnKF,~\cite{sakov_iterative_2012}) and smoother (IEnKS,~\cite{bocquet_iterative_2014}) have been shown to clearly outperform the standard EnKF and EnKS for certain dynamical systems. IEnKF has also been extended in order to take into account the model errors in~\cite{sakov_iterative_2018}. The recently proposed quantile-conserving ensemble filter framework~\citep{anderson_quantile-conserving_2022} supports arbitrary distributions for the prior and observation likelihood. Detailed descriptions of the ensemble Kalman smoother can be found in e.g.~\cite{cosme_smoothing_2012, bannister_review_2017, bergou_convergence_2019}.

\subsection{Extensions beyond the standard data assimilation setup}
\label{sec:da_extensions}
The ``standard'' DA scenario in the context of Eq.~\eqref{eq:state_equation}-\eqref{eq:Gaussianbackgroundprior} applies to many important problems and provides a useful framework for defining and comparing the major classes of DA methods. However, as ADDA supports assimilation tasks beyond this standard scenario, we next describe its capabilities in full generality.

ADDA supports a more general class of dynamical systems than the nonlinear dynamics with Gaussian noise of Eq.~\eqref{eq:state_equation}, including both deterministically and probabilistically evolving systems with arbitrary noise distributions as in Eq.~\eqref{eq:prior_factorisation}. ADDA's strong-constraint 4D-Var and Ensemble Kalman methods require only a (possibly stochastic) time stepping function~$\mathcal M$. The process noise distribution is restricted to be Gaussian for (ensemble) Kalman methods but can take any form for weak-constraint 4D-Var, as long as the associated log-probability can be computed analytically. In addition, while the time-discretized dynamics in Eq.~\eqref{eq:state_equation} imply a fixed time step, ADDA also supports systems evolving continuously in time, for which a time step~$dt$ controls the extent of time integration on each call to~$\mathcal M$.
This provides support for irregularly spaced observations, as demonstrated in Sect.~\ref{sec:irregular_time_example}. Furthermore, while Eq.~\eqref{eq:state_equation} assumes an autonomous system, in ADDA~$\mathcal M$ can receive external inputs such as boundary conditions or forcings. An illustrative example of assimilating a system with a forcing input will be considered in Sect.~\ref{sec:4dvargotm}.

ADDA also supports a more diverse set of observation types. Eq.~\eqref{eq:observation_equation} assumes Gaussian observation noise~$\boldsymbol{\eta}_t$ around a deterministically computed conditional mean $\mathcal H_t(\mathbf{x}_t)$, but ADDA supports arbitrary conditional distributions of observations given states $p(\mathbf{y}_t|\mathbf{x}_t)$. This allows for heteroscedastic and non-Gaussian observations, which can occur in atmospheric, ocean and ecosystem dynamics. It also supports observations that depend on the state at multiple time points, for example through temporal interpolation or direct observation of the slope for a long-term trend.

\section{Implemented structures and methods} \label{sec:methods}
Here, we give an overview of DA methods implemented in ADDA: the Kalman filter and smoother, ensemble Kalman filter and smoother, and strong- and weak-constraint 4D-Var with a single or a sliding window. We also give some technical details on the expected structure of the state and observation variables, which are implemented in a general enough way to allow for easy extensions to numerous other assimilation methods. We do not specifically discuss the implementations of the Kalman filter and smoother, as those are simply consistent with standard practice.

\subsection{The state variable}
\label{sec:state}

The state~$\mathbf{x}_t$ of a dynamical system, evolved in time by Eq.~\eqref{eq:state_equation}, is implemented in ADDA's \verb+State+ class, designed for maximum flexibility and used by all DA methods. A \verb+State+ instance represents the system state at one or more time steps, i.e. it contains~$\mathbf{x}_{0:T}$, along with time stamps~$\tau_{0:T}$ for each step. This design choice allows for representation of state sequences with variable time stepping.
ADDA's \verb+State+ class builds on the PyTorch \verb+TensorDict+ class~\citep{bou_torchrl_2024}, which encapsulates multiple \verb+torch.Tensor+ arrays with support for parallelization across shared dimensions. \verb+State+ objects incorporate two shared dimensions: a batch/ensemble dimension and the time axis. The resulting parallelization of time stepping across ensemble members and time steps is critical for efficient implementation of certain DA methods.

In addition to these shared dimensions, the system state can incorporate multiple named fields, each with its own dimensionality and spatial structure. Thus, system states can range from simple unstructured vectors to complex structures incorporating many variables on staggered or collocated grids, or unstructured meshes. This ability to incorporate variable fields at multiple resolutions is used for the two-level Lorenz system~\citep{lorenz_designing_2005}, with a practical example in Sect.~\ref{sec:TTL96}.


\subsection{Observation operators}
\label{sec:obs_op}

A central element of data assimilation is the observation operator, which relates the state~$\mathbf{x}$ to the observations~$\mathbf{y}$ through Eq.~\eqref{eq:observation_equation} (for Gaussian observation noise), Eq.~\eqref{eq:obs_error_factorisation} (other distributions, one time point per observation) or Eq.~\eqref{eq:bayes} (full generality). In ADDA, the observation operator is a class implementing methods for sampling, evaluating and generally working with the density $p(\mathbf{y}_{0:T}|\mathbf{x}_{0:T})$.

The observation operator's density evaluation method must support automatic differentiation for use with variational methods, but this is not required for (ensemble) Kalman methods. Observation operators can also optionally implement the conditional mean~$\mathcal H$ for use by Ensemble Kalman methods. For Kalman filtering and smoothing, the conditional mean must also be a linear function of the state.

While ADDA users can easily define their own classes inheriting from a base \verb+ObservationOperator+, implementations are also provided for several common observation scenarios. The simplest of these describes noisy observations of all state variables, with the option for noise levels that vary across time, location or field. Another pre-built observation operator supports masking, with observations for only some fields, locations or time steps. ADDA also provides operators describing arbitrary linear transformations from states to observations for use with (ensemble) Kalman methods, as well as temporal interpolation of state variables between time steps.

\subsection{System dynamics}
\label{sec:dynamics}
ADDA implements the system dynamics~$\mathcal M$ as a function (python \verb+Callable+ type) with four inputs: $\mathbf{x}_t$, $dt$, $F^d_t$ and $F^s$. The time step~$dt$ controls how far in time the system should be advanced (possibly using smaller or adaptive internal steps). Dynamic inputs~$F^d_t$ are provided at each time step to represent time-varying forcings or boundary conditions, while static inputs~$F^s$ represent configurable aspects of the system that do not change over time, such as scalar parameters (e.g. fluid viscosity) or spatially extended fields (e.g. ocean bathymetry). While~$\mathcal M$ always takes 4 inputs to ensure compatibility across ADDA's interfaces, in many practical cases, not all 4 inputs will be used in a specific system implementation. We present in Sect.~\ref{sec:irregular_time_example} an experiment where specifying the time increment~$dt$ is necessary, and in Sect.~\ref{sec:4dvargotm} an example where dynamic inputs are provided.

\subsection{Heuristics for initializing the states}
For both variational and ensemble methods, the initialization of system states can impact both the accuracy and speed of DA algorithms. For variational methods, the assimilation cost is usually not convex, so different initializations might lead to different local minima of the cost function. Ensemble methods require an initial state ensemble, and may never converge to sensible posterior estimates of the state distribution if the initial estimate is too far from the true initial state.

In some cases, system states can be initialized using historical averages such as Earth system climatologies. When these are unavailable or insufficiently precise, ADDA offers a flexible set of initialization heuristics that make use of available observations. Examples of these initializations include linear interpolation of observations over time, nearest neighbor interpolation, and user-provided default values for each field and/or location.

\subsection{Variational methods}

\subsubsection{Strong-constraint 4D-Var}
Strong-constraint 4D-Var, often simply called 4D-Var, is a widely used variational DA method (technical description in Sect.~\ref{sec:4dvar}). While deriving an adjoint model for complex dynamical systems is notoriously difficult, computing the necessary adjoints is straightforward in an automatic differentiation framework. 
In the ADDA implementation, the user is asked to provide the dynamical model~$\mathcal{M}$ following the conventions described in Sect.~\ref{sec:dynamics}, the set of observations~$\mathbf{y}_{0:T}$ and the corresponding observation operator as described in Sect.~\ref{sec:obs_op}. 
The prior distribution on $p(\mathbf{x}_0)$ can be provided, but will otherwise be assumed to be uniform so the first term of the right-hand side vanishes in Eq.~\eqref{eq:sc4dvar}. Forcing and/or boundary data can also be provided for each time step.


Strong-constraint 4D-Var usually assumes Gaussian distributions for the initial state prior and observation noise, but ADDA's implementation does not require these assumptions. While the examples in Sect.~\ref{sec:examples} use Gaussian distributions, ADDA users can define and employ other distribution classes implementing density evaluation and sampling methods.

\subsubsection{Weak-constraint 4D-Var}
\label{sec:wc4dvar}

ADDA computes the loss in Eq.~\eqref{eq:wc4dvar} using parallelization over the time axis for $\mathcal{M}(\mathbf{x}_{0:T-1})$. This parallelization considerably reduces the overall computation time for each optimization step, especially on GPUs.

In addition to the inputs required for strong-constraint 4D-Var, the weak-constraint variant requires a model error distribution for all time steps on which the optimization is performed. 
ADDA does not require model error distributions to be Gaussian.

\subsubsection{Sliding-window 4D-Var}
\label{sec:sliding_window}
ADDA's variational DA methods can be used with a sliding-window approach, in which observations are divided into assimilation sub-windows, and a separate 4D-Var problem is solved for each sub-window. The result of each sub-window's optimization is used to define the mean of a background prior for the subsequent sub-window, by directly selecting (weak-constraint) or simulating to (strong-constraint) the appropriate time step. ADDA's sliding-window 4D-Var methods further support a \verb+covariance_factor+ argument that controls the tightening of the background prior for all but the first assimilation sub-window, reflecting the fact that a previously assimilated sub-window decreased the uncertainty on the initial state for the next sub-window. As illustrated in Sect.~\ref{sec:QG}, a covariance factor $<1$ can be used to express increased confidence in the prior distribution when some observations have been assimilated in a previous sub-window. ADDA supports both overlapping and non-overlapping assimilation sub-windows.

This approach allows variational DA to be applied to long observation sequences, for which optimization over a single window would encounter memory constraints, or for which the growing input-output gradients of chaotic dynamical systems would destabilize the optimization. Our implementations of sliding-window strong- and weak-constraint 4D-Var use essentially the same arguments as their single-window versions, with additional arguments for the sizes of the sub-windows and of the shifts between them, as well as construction of background priors after the first sub-window. 

\subsubsection{Gradient descent for variational methods}
\label{sec:optimizer}
ADDA's variational DA methods employ gradient-based optimization. While computing gradients through $\mathcal M$ by automatic differentiation, we leverage the extensive tools for differentiable optimization implemented in PyTorch. Variational methods support PyTorch built-in as well as user-defined \verb|Optimizer| classes, with support for \verb|LRScheduler| classes for finer control of the optimization process. Additional arguments controlling optimizer behavior (e.g. line search function or momentum weight) can be passed to variational DA methods. The default optimizer is the limited-memory BFGS~\citep{liu_limited_1989}.

\subsection{Ensemble methods}
An advantage of ensemble-based DA methods is that they do not require gradients from an adjoint model of the dynamics~$\mathcal{M}$. Thus, ADDA's ensemble methods do not employ automatic differentiation, but nonetheless benefit from efficient GPU-accelerated computation in PyTorch, with parallelization across the ensemble to accelerate DA for large ensembles or states. 
Our implementation follows the classical ``perturbed observations'' approach, introduced by~\cite{burgers_analysis_1998}.

ADDA implements EnKF as a special case of the EnKS with zero lag, resulting in a simple and compact codebase. When employing an EnKS method, key choices include:
\begin{itemize}
    \item The distribution from which to sample the initial ensemble (usually a Gaussian)
    \item The size of the ensemble, i.e. its number of members
    \item The lag over which to perform the smoothing
    \item The choice of linearizing the observation operator or using a nonlinear expression
\end{itemize}
All of these choices can be easily made in ADDA by specifying arguments when calling the EnKF or EnKS functions. In our implementation, the distribution of the initial ensemble is a diagonal Gaussian. Importantly, while some variants of the EnKS algorithm model the full smoothing distribution $p(\mathbf{x}_t|\mathbf{y}_{0:T})$, ADDA's EnKS is a fixed-lag smoother which makes the common assumption that the posterior on states at a given time step~$t$ is only influenced by the observations that are located at most~$d$ time steps later than~$t$, where~$d$ is a chosen delay parameter. However, one can simply define $d \geq T$ in order to obtain the full smoothing distribution.

ADDA's default behavior follows most EnKF/S methods in linearizing the conditional mean observation~$\mathcal{H}$ as a function of the state, but 
ADDA also supports the method described by e.g.~\cite{bannister_review_2017} that instead applies a nonlinear $\mathcal H$ to each ensemble member. Linearization requires the observation operator's class implement a \verb|linearize| method, and all included observation operators (Sect.~\ref{sec:obs_op}) do so. Note that the resulting matrix~$\mathbf{H}$ has a size of $(s, o_t)$. When both of these numbers are high, this can represent a very large matrix. Yet, in some cases, such as the ``binary masking'' observation operator, the structure of this matrix is extremely sparse by design. Thus, in order to avoid unnecessary computations, our implementations use sparse data structures from PyTorch in such cases.

\subsection{Specification of the state time axis}
ADDA's structure is centered on the \verb+State+ object (Sect.~\ref{sec:state}), whose attributes include the \verb|TensorDict| class \verb|fields| and the \verb|Tensor| class \verb|time_axis|. One important consideration when solving a data assimilation problem is the definition of the state time axis for the inferred trajectory. In the simplest cases, the inferred state will be defined on the same time axis as the set of available observations, which is the default behavior of our package. However, in some cases, the state variable might be defined with a much finer time resolution than the observations for numerical stability reasons. Conversely, the observations might be defined at a very fine temporal resolution, so that inferring a state trajectory with this resolution would be unnecessarily costly. Observations may also be defined with irregular time steps. 
For these reasons, ADDA enables defining a time axis for inferring a state trajectory in 4D-Var, independently of the observation time axis. In this case, the observation operators must define the mapping from state trajectories to distributions on observation sets, defined on different time axes. ADDA currently provides one such operator that involves linear interpolation of a state trajectory with a regular time axis on an arbitrary (possibly irregular) time axis. An example illustrating this feature is presented in Sect.~\ref{sec:irregular_time_example}. ADDA's ensemble-based methods require the states and observations to use the same time axis, but allow uneven irregular spacing along this axis as shown in Sect.~\ref{sec:irregular_time_example}. ADDA's weak-constraint 4D-Var supports irregularly spaced observations along the time axis but requires regularly spaced system states to avoid complicating parallel computations.

\section{Illustrative experiments}
\label{sec:examples}

We demonstrate ADDA's capabilities on multiple experiments, covering a wide variety of dynamical systems, observation operators and DA techniques. As previously stated, variational DA requires dynamical systems to support automatic differentiation, with an implementation in either PyTorch or JAX. We used our own PyTorch differentiable implementations of the dynamical systems for most of the examples associated with the package, including linear dynamics, Lorenz-63 (L63), Lorenz-96 with one (L96-1L) and two (L96-2L) timescales, Kuramoto-Sivashinsky (KS-1D), Korteweg–De Vries (KdV), Kolmogorov flow\footnote{For this particular system, we use a neural emulator rather than a differentiable simulation.} (KF) and advection and diffusion operators with an inert tracer. For the three-layer quasi-geostrophic system (QG-3L), we use the PyTorch implementation from~\cite{thiry_modified_2023} in Sect.~\ref{sec:QG}. For the two-dimensional Kuramoto-Sivashinsky system (KS-2D), we use the Exponax library~\citep{koehler_apebench_2024} built in JAX, in order to demonstrate the ability of ADDA to bridge PyTorch and JAX gradients, in Sect.~\ref{sec:ksjax4dvar}. We also used a JAX-PyTorch bridge with the Exponax implementation of the KdV system, in addition to our own PyTorch implementation.
Our codebase includes multiple notebooks to illustrate the ease with which ADDA can be applied to these systems with diverse observation operators, without requiring the user to implement extensive custom software routines for each case. In the following sections, we present a selected subset of these examples in greater detail.

\subsection{An introduction to variational data assimilation on the Lorenz-63 system}

Our first example illustrates strong- and weak-constraint 4D-Var on the Lorenz-63 dynamical system, described in~\cite{sparrow_lorenz_1982}. The Lorenz-63 system is a major testbed for data assimilation. It is a very simple and low-dimensional set of equations, yet its chaotic behavior makes long-term forecasting impossible. The system has 3 variables $x, y, z$, from which the respective time derivatives can be written as follows:
\begin{equation}
\label{eq:L63}
    \dot{x} = \sigma(y-x) ; \quad \dot{y} = x(r-z)-y ; \quad \dot{z} = xy-bz.
\end{equation}
The parameters of these equations are set to their most common values $(\sigma, r, b) = (10,28,8/3)$. With these settings, the system's Lyapunov time (for which small perturbations grow by a factor of~$e$) is roughly~1.10.
As a reminder, we will use the notation $\mathbf{x} = (x, y, z) \in \mathbb{R}^3$ for the full state of the system, not to be mistaken with $x \in \mathbb{R}$, the first variable of the system. Conversely,~$\mathbf{y}$ (not to be mistaken with~$y$) denotes the set of available observations. The groundtruth time series for all experiments on this system are obtained by integrating the system~\eqref{eq:L63} with a time step of~$0.01$, using the 4th order Runge-Kutta method.

\subsubsection{Assimilating noisy but complete observations}

\begin{figure}
    \centering
    \includegraphics[width=\linewidth]{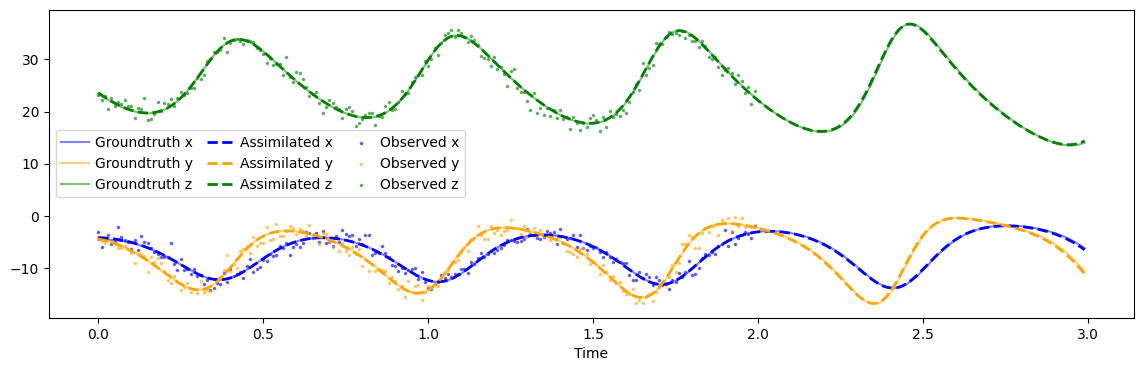}
    \caption{Visualization of the true values, noisy observations and assimilated trajectory for the Lorenz-63 system, with strong-constraint 4D-Var, in a case where all variables are observed with a Gaussian noise until time 2.0.}
    \label{fig:L63_sc4dvar}
\end{figure}

We first consider an observation operator that adds Gaussian noise:
\begin{equation}
    \mathbf{y}_t = \mathbf{x}_t + \boldsymbol{\eta}_t,
\end{equation}
where $\boldsymbol{\eta}_t \sim \mathcal{N}(\mathbf{0}, \mathbf{I}_3)$, and $\mathbf{I}_3$ is the $3 \times 3$ identity matrix, so that the observation noise is uncorrelated. In this case, the assimilation task reduces to a denoising task. We solve it using strong-constraint 4D-Var over a window of~$2$ time units, i.e., $200$ time steps with no background prior term (implicitly corresponding to a uniform background prior on~$\mathbb{R}^3$). We initialize the variables with the observed values at the initial time and perform~5 L-BFGS optimization steps with a learning rate of~$0.1$.  The result is an assimilated initial state, which is expected to be significantly closer to the true initial state than the initial guess (i.e., the noisy observations at the first time step). This estimated initial state is then rolled out in time using the same numerical integration scheme as for generating the groundtruth time series. As can be seen from Fig.~\ref{fig:L63_sc4dvar}, we can convincingly denoise the observations and even accurately extend the predictions beyond the range of the observations.

\subsubsection{Imputing a missing system state variable}
We next consider a case where the variables~$x$ and~$y$ are always observed with a Gaussian noise of standard deviation~$1$, while the variable~$z$ is never observed:
\begin{equation}
    \mathbf{y}_t = \mathcal{H}(\mathbf{x}_t) + \boldsymbol{\eta}_t = (x_t, y_t) + \boldsymbol{\eta}_t, \quad \boldsymbol{\eta}_t \sim \mathcal{N}(0, \mathbf{I}_2)
\end{equation}
where $\mathcal{H}: \mathbb{R}^3 \to \mathbb{R}^2$ truncates the third variable: $\mathcal{H}(x, y, z) = (x, y)$. A key goal of the assimilation is to infer the missing variable~$z$ over time by using only the available (noisy) observations on~$x$ and~$y$. Since the observation operator~$\mathcal{H}$ can be differentiated using straightforward automatic differentiation, strong-constraint 4D-Var can be applied as in the denoising example. The only difficulty is initializing the unobserved variable~$z$ before starting the gradient descent. We arbitrarily set the initial value to~$25.0$, which is relatively close to the stationary mean of this variable.

\begin{figure}
    \centering
    \includegraphics[width=\linewidth]{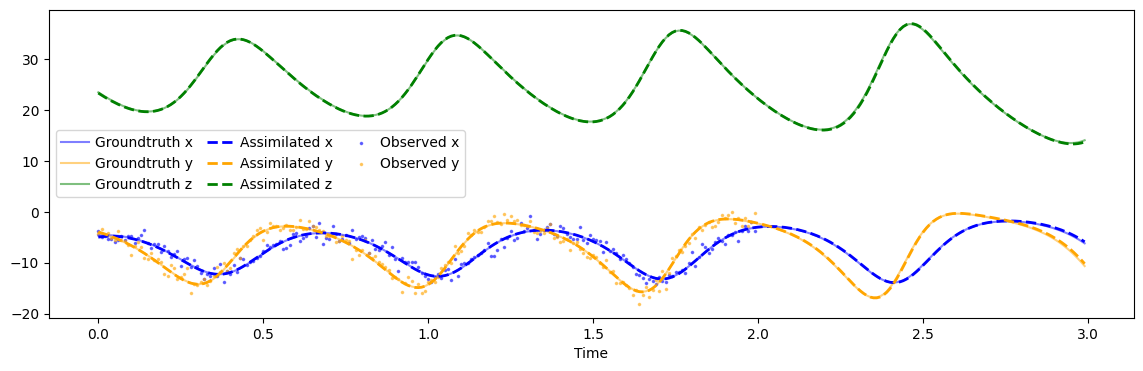}
    \caption{Visualization of the true values, noisy observations and assimilated trajectory for the Lorenz-63 system, with strong-constraint 4D-Var, in a case where observations are entirely missing for the~$z$ variable.}
    \label{fig:L63_sc4dvar_missing}
\end{figure}

As can be seen in Fig.~\ref{fig:L63_sc4dvar_missing}, despite the more difficult observation setting, we can accurately reconstruct the true state, including the unobserved third variable. Extending the assimilation beyond the window of observations again results in accurate forecasts.

\subsubsection{Weak-constraint 4D-Var on a long window}
\label{sec:L63_wc4dvar}

Single-window strong-constraint 4D-Var limits the duration of observation sequences for chaotic systems such as Lorenz-63, as chaotic behavior makes it impossible to optimize initial conditions based on observations far into the future. In practice, this results in ill-behaved gradients when the assimilation window is long relative to the Lyapunov time of a system, preventing the method from converging to the correct solution, even with a favorable observation pattern. In contrast, weak-constraint 4D-Var does not require the optimized state trajectory to precisely follow the known system dynamics $\mathcal M$, and its long-term predictions need not arise deterministically from a single initial state. Instead, our implementation of weak-constraint 4D-Var relies only on one-time-step predictions, as detailed in Sect.~\ref{sec:wc4dvar}. This enables both numerical stability and fast parallel computation. To demonstrate this, we now consider a much longer observation window with~100 time units, corresponding to~$10^4$ time steps. Observations are randomly masked, and each variable at each time step is observed with~${25\,\%}$ probability. Observations are corrupted by additive Gaussian white noise with standard deviation~2, higher than in the previous examples. Note that this strategy does not result in regularly spaced observations for any of the~3 variables.

\begin{figure}
    \centering
    \includegraphics[width=\linewidth]{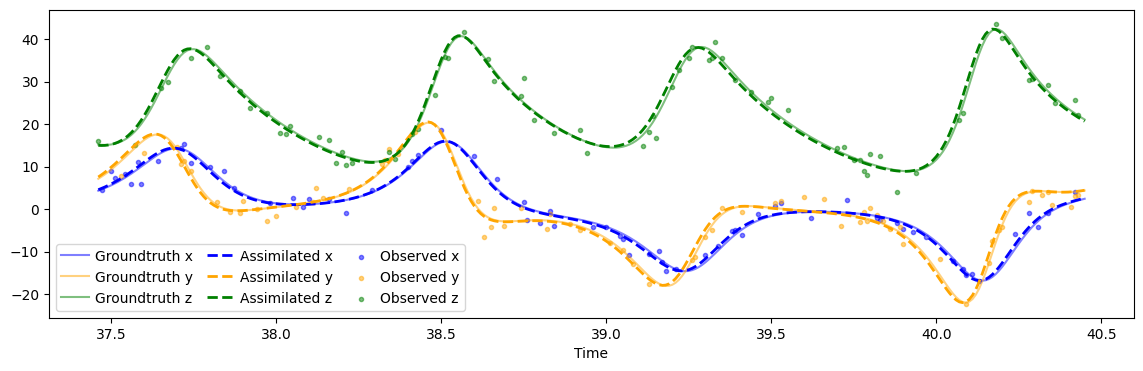}
    \caption{Visualization of the true values, noisy observations and assimilated trajectory for the Lorenz-63 system, with weak-constraint 4D-Var performed on a long time series with a sparse observation pattern.}
    \label{fig:L63_wc4dvar}
\end{figure}

We use weak-constraint 4D-Var with no background prior term and a diagonal Gaussian with mean zero and standard deviation $10^{-2}\mathbf{I}_2$ for the model error distribution. As weak-constraint 4D-Var optimizes over the full system state trajectory, we initialize this trajectory using nearest-neighbors interpolation over time for each variable. We optimize using~50 L-BFGS iterations, with a learning rate of~1. We display the assimilation result for a randomly selected crop of the observation window in Fig.~\ref{fig:L63_wc4dvar}. Despite the relatively low signal-to-noise ratio and the sparsity of the observations, the assimilated trajectory closely matches the true trajectory.

\subsection{Data assimilation on the Lorenz-96 system}
\label{sec:L96}

The Lorenz-96 system~\citep{lorenz_optimal_1998} is another relatively simple chaotic system, designed to represent the evolution of an atmospheric variable over a latitude band. It is more complex than Lorenz-63 due to its spatial structure: the system has~$n$ variables $x_1, ..., x_n$ on a periodic domain, with their time derivatives expressed as:
\begin{equation}
\label{eq:L96}
    \dot x_i = (x_{i+1} - x_{i-2}) x_{i-1} - x_i + F.
\end{equation}
Time is expressed in arbitrary units with one unit corresponding to~$5$ days of the evolution of the atmosphere, in terms of Lyapunov time and predictability.
We use the most common setup, with $n=40$ and $F=8$, for which the Lyapunov time is approximately~0.60. In Sect.~\ref{sec:L96_4dvar}, and~\ref{sec:L96_EnKFS}, the equations are integrated numerically using a 4th-order Runge-Kutta scheme with a time step of~0.01, yielding weakly nonlinear dynamics at each time step. In Sect.~\ref{sec:irregular_time_example}, they are integrated with an irregular time step, as specified later.

The experiments of Sect.~\ref{sec:L96_4dvar},~\ref{sec:L96_EnKFS} and~\ref{sec:irregular_time_example} all use the same observation setup, consisting of a binary mask and white Gaussian noise. At each time step, we randomly mask~30 of the~40 variables. For the remaining variables, we add a Gaussian noise of variance~$1$.

\subsubsection{Strong- and weak-constraint 4D-Var}
\label{sec:L96_4dvar}

The 4D-Var algorithms are initialized with nearest-neighbor interpolation over time, as in Sect.~\ref{sec:L63_wc4dvar}. 
For the strong-constraint variant, only the estimate of the initial state is required. For the weak-constraint variant, we consider the model errors to follow a Gaussian distribution with zero mean and a diagonal covariance matrix with diagonal coefficients~$10^{-4}$, as in Sect.~\ref{sec:L63_wc4dvar}.
Due to the chaotic nature of the Lorenz-96 system, we restrict the strong-constraint optimization to an observation window of~100 time steps to maintain stability, while the weak-constraint assimilation is performed on a window of~800 time steps. The assimilation costs of both 4D-Var variants are minimized using L-BFGS, with a learning rate of~$0.1$ for strong-constraint and~$1$ for weak-constraint.

\begin{figure}
    \centering
    \includegraphics[width=\linewidth]{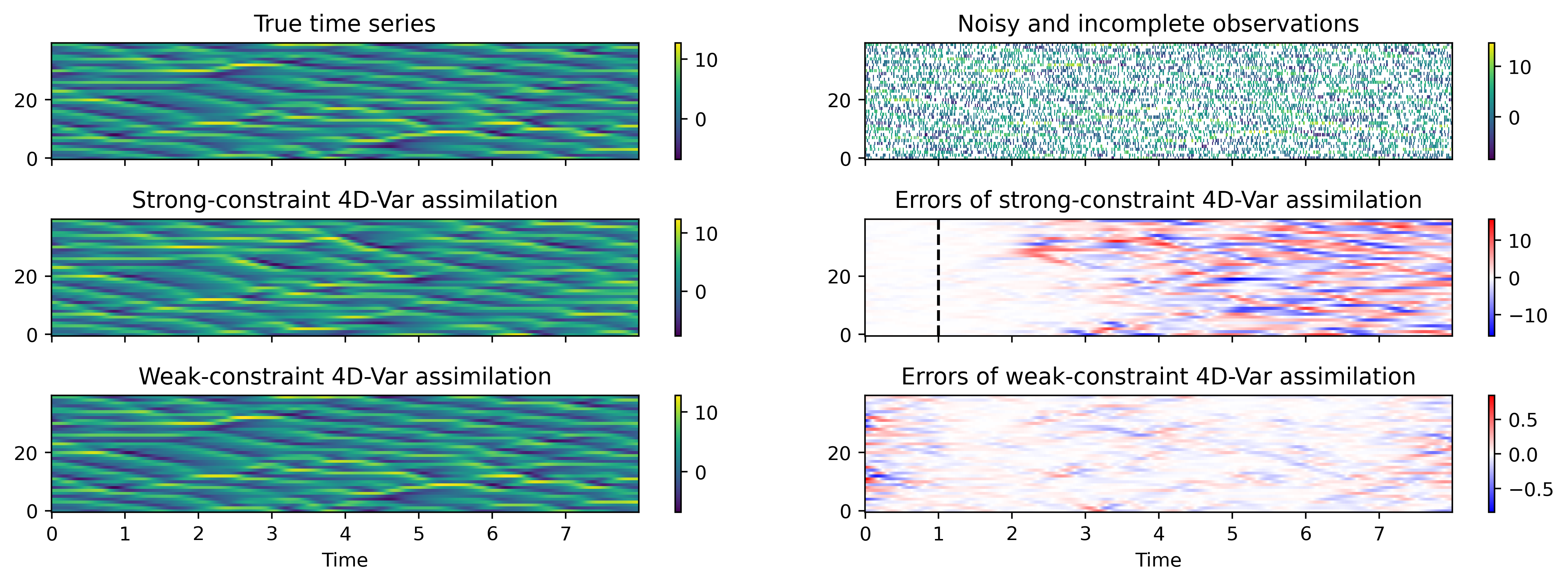}
    \caption{Visualization of the true state values, the corresponding sparse and noisy observations, and the state estimations and errors obtained by strong- and weak-constraint 4D-Var on the Lorenz-96 dynamical system. The dashed vertical line marks the limit of assimilated observations for strong-constraint 4D-Var.}
    \label{fig:L96_4dvar}
\end{figure}

Results, presented in Fig.~\ref{fig:L96_4dvar}, are consistent with the expected behavior of the methods; the absolute error at the beginning of the time series with respect to the true state is significantly smaller than the amplitude of the observation noise. Due to the chaotic nature of the Lorenz-96 system, the absolute error of strong-constraint 4D-Var grows exponentially once the integration extends beyond its observation window. Thus, after roughly~4 time units (i.e., 400 time steps), the forecast becomes entirely unskilled, in the sense that the estimated state is only as close to the true state of the system as the expected distance between two independent states sampled from the stationary distribution of the Lorenz-96 system. This could be alleviated by periodically performing a new assimilation with new observations, corresponding to sliding-window 4D-Var. 

In contrast to single-window strong-constraint 4D-Var, weak-constraint 4D-Var can successfully assimilate all observations in a single window spanning several Lyapunov times of the system, which is made possible by the introduction of model errors. As a result, assimilation errors are relatively low and stable in time. The largest errors are observed on the boundaries of the assimilation window, particularly on the left boundary, a common phenomenon explained by the relative scarcity of surrounding observations at those time steps.

\subsubsection{Ensemble Kalman filter and smoother}
\label{sec:L96_EnKFS}


We next apply the Ensemble Kalman Filter (EnKF) and Smoother (EnKS) to the Lorenz-96 system, with the same observation scenario as previously.
Key factors influencing the success of ensemble-based methods include the ensemble size and initial state distribution. In many works involving variants of EnKF for the Lorenz-96 system (e.g.~\cite{sakov_iterative_2012, bocquet_joint_2013}), knowledge of the true initial state is used to define the ensemble mean. While this sort of experiment can reflect a realistic setup in which a previous observation window provides an informative prior, we found it essential for ADDA to incorporate and test initializations relying solely on available observations, as for variational methods. This choice is crucial since the quality of the initialization of EnKF and EnKS largely affects their capacity to converge to values close to the true state.

We therefore initialize EnKF/S using nearest-neighbor interpolation over time, as previously. With this heuristic, the error of the initial guess relative to the true initial state is expected to exceed the observation noise, since it is driven by two factors: the observation noise itself and the fact that the observed value comes from a subsequent time step. We neglect this second factor and set the standard deviation of the initial ensemble to~$1.0$, the same as the observation noise. With such an initialization, a relatively high ensemble size was required to prevent divergence as the ensemble methods proceeded over subsequent time steps. 
A more accurate initial ensemble mean would enable a significant reduction in the ensemble size, as thoroughly analyzed in e.g.~\cite{bocquet_online_2021}. 
We choose an ensemble size of~50 for our main results, which ensures fast convergence and an accurate estimate of the system state. We also use an inflation factor of~$1.01$. For the EnKS, we use a lag of $d=50$ time steps.

\begin{figure}
    \centering
    \includegraphics[width=\linewidth]{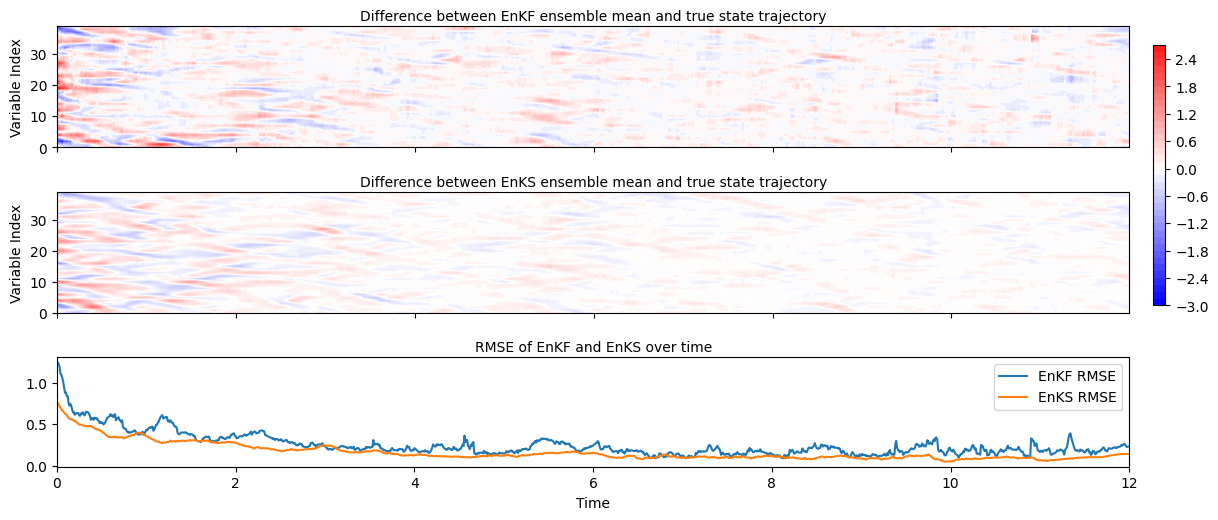}
    \caption{Comparison between the errors obtained by the assimilations of an Ensemble Kalman Filter (EnKF) and a similarly parameterized Ensemble Kalman Smoother (EnKS) on the same set of observations from the Lorenz-96 dynamical system.}
    \label{fig:L96_EnKF_EnKS}
\end{figure}

The results of the EnKF and EnKS assimilations are shown in Fig.~\ref{fig:L96_EnKF_EnKS}. The largest errors for both methods occur in the first few time steps of the assimilation window. These large errors are caused by the imperfect initialization described above, after which the ensemble mean progressively gets closer to the true state. The figure also clearly shows the benefit of the smoothing over filtering, as the root mean squared error from the former is significantly lower throughout the assimilation window.

\subsubsection{Handling observations with an irregular time sampling pattern}
\label{sec:irregular_time_example}

In the previous examples, the observations are sampled from a regular temporal grid, with a fixed time step. Here, we show that our framework is substantially more flexible, as it imposes no constraint on the regularity of the time steps of the observations.

Regarding strong-constraint 4D-Var and the ensemble Kalman methods, the irregular temporal sampling does not induce additional difficulty. By default, these methods estimate the state on exactly the timestamps of the observations. For the ensemble Kalman methods, this is in fact the only available mode in ADDA, as these methods do not have a \verb|state_time_axis| argument. In contrast, the 4D-Var methods allow the state time axis to be specified independently from the observation time axis, with the constraint that the state time axis must be regular for weak-constraint 4D-Var (to enable the parallel computations). When the state time axis differs from the observation time axis (which, in this section, applies only to weak-constraint 4D-Var), the observation operator should be able to relate them. As mentioned in Sect.~\ref{sec:obs_op}, we currently provide an observation operator class that linearly interpolates a regularly sampled state time axis to the observation time axis. 

We uniformly sample~$800$ observation times over a time domain of~$8$ time units and sort them, so that the mean difference between two consecutive observation times is~$0.01$, matching our other L96 experiments. The groundtruth time series is generated by integrating Eq.~\eqref{eq:L96} on precisely this irregular time axis.

\begin{figure}
    \centering
    \includegraphics[width=1.0\linewidth]{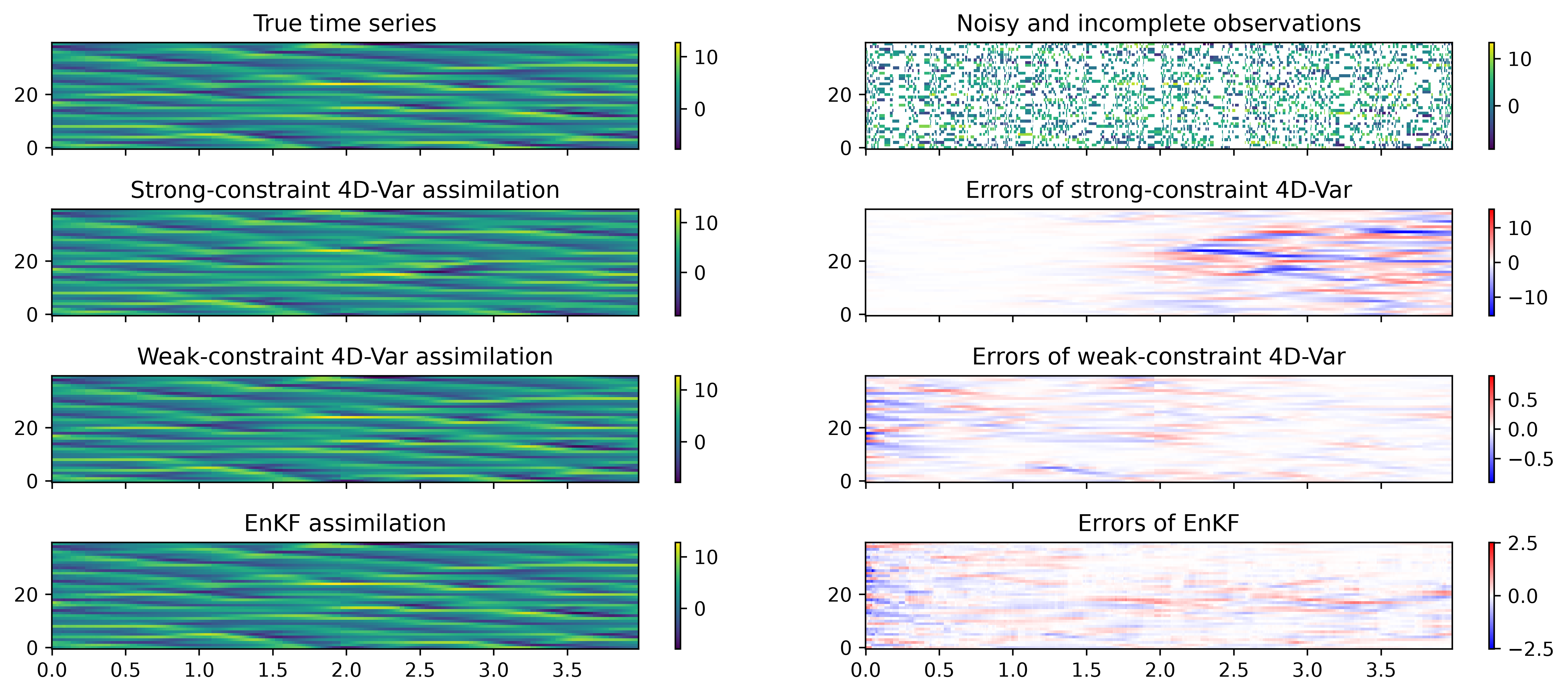}
    \caption{Visualization of the groundtruth time series for the Lorenz-96 dynamics with irregular time sampling, the corresponding sparse and noisy observations and the assimilated states and associated errors for the strong- and weak-constraint 4D-Var and EnKF. For ease of visualization, we only show the first half of the whole observation domain.}
    \label{fig:irregular_time_L96}
\end{figure}

We show the groundtruth, observations and assimilation results of all methods in Fig.~\ref{fig:irregular_time_L96}. These results are largely consistent with those from the regular sampling case in the previous sections. Indeed, the single-window strong-constraint 4D-Var method can accurately match the true state over the first few time units before it diverges due to the chaotic nature of the system, while the weak-constraint 4D-Var and EnKF methods accurately reconstruct the true time series, with the highest errors on the edges of the observation window for weak-constraint 4D-Var and at the beginning of the window for EnKF. None of these data assimilation methods appears to be significantly affected by the irregular time axis of the observations. Weak-constraint 4D-Var obtains, as in the previous experiments, the lowest assimilation errors.

\subsection{Weak-constraint 4D-Var on the two-timescale Lorenz-96 system}
\label{sec:TTL96}

Following the examples on the basic Lorenz-96 system described in Sect.~\ref{sec:L96}, we now consider the two-timescale extension of this system, discussed in e.g.~\cite{lorenz_optimal_1998} and~\cite{lorenz_designing_2005}. The system consists of two sets of variables: $n$ slow-evolving variables $(X_k)_{k=1}^n$ and, for each of these, $J$ fast-evolving variables $Y_{1,k}, ..., Y_{J,k}$. Thus, the system comprises~$n$ slow variables and~$nJ$ fast variables, for a total dimension of~$n(J+1)$. These variables evolve in time according to:
\begin{align}
    &\label{eq:2TS_L96_slow}
    \dot X_k = (X_{k+1} - X_{k-2}) X_{k-1} - X_k + F - \frac{hc}{b} \sum_{j=0}^J Y_{j,k}, \\
    &\label{eq:2TS_L96_fast}
    \dot Y_{j,k} = - cbY_{j+1,k} \left( Y_{j+2,k} - Y_{j-1,k} \right) - c Y_{j,k} + \frac{hc}{b} X_k.
\end{align}

As shown by Eq.~\eqref{eq:2TS_L96_slow}, the slow variables $X_1, ..., X_n$ evolve similarly to the state of the single-timescale Lorenz-96 system in Eq.~\eqref{eq:L96}, except that they are also coupled to the fast variables through the last term of the equation. The fast variables~$Y_{j,k}$ are influenced by their neighbors and their associated slow variables, and typically evolve more rapidly. 
$F$,~$b$,~$c$ and~$h$ respectively characterize the forcing on the slow variables, the mutual influence between the fast variables, the relative rate of evolution of the fast variables compared to the slow ones and the strength of the coupling between slow and fast variables. We numerically integrate Eq.~\eqref{eq:2TS_L96_slow} and~\eqref{eq:2TS_L96_fast} using the most common parameter values: $h=1, b = c = F = 10$. We set the size of the system to $n=36$ and $J=10$, yielding $n=36$ slow variables and $nJ=360$ fast variables. We use the 4th order Runge-Kutta scheme, with a time step of~0.01 and a 1000-step assimilation window.

The two-timescale Lorenz-96 system is challenging to incorporate in DA systems that require a fixed spatial grid for the system states, since here the state consists of two periodic fields of different sizes, one of them being one-dimensional and the other two-dimensional. This state can, however, be conveniently represented in ADDA with a single \verb|State| object, using the structure described in Sect.~\ref{sec:state}. Likewise, a binary masking observation operator with different observation proportions and noise standard deviations for the two fields can be defined straightforwardly by defining the observation proportion and noise amplitudes as dictionaries containing one floating value for each field. While still a simple system, this illustrates how ADDA handles systems with heterogeneous state structures, which is necessary to accommodate realistic geophysical problems.

We apply weak-constraint 4D-Var in~2 observation scenarios. In Sect.~\ref{sec:TTL96-1}, sparse and noisy observations are available for both the slow and the fast variables. In Sect.~\ref{sec:TTL96-2} and~\ref{sec:TTL96-3}, we have access to sparse and noisy observations of only the slow variables. The fast variables are ignored by the DA procedure in Sect.~\ref{sec:TTL96-2} (effectively acting as unresolved sub-grid-scale processes) but explicitly inferred in Sect.~\ref{sec:TTL96-3}. 

\subsubsection{Assimilation of jointly observed slow and fast variables}
\label{sec:TTL96-1}

\begin{figure}
    \centering
    \includegraphics[width=1.0\linewidth]{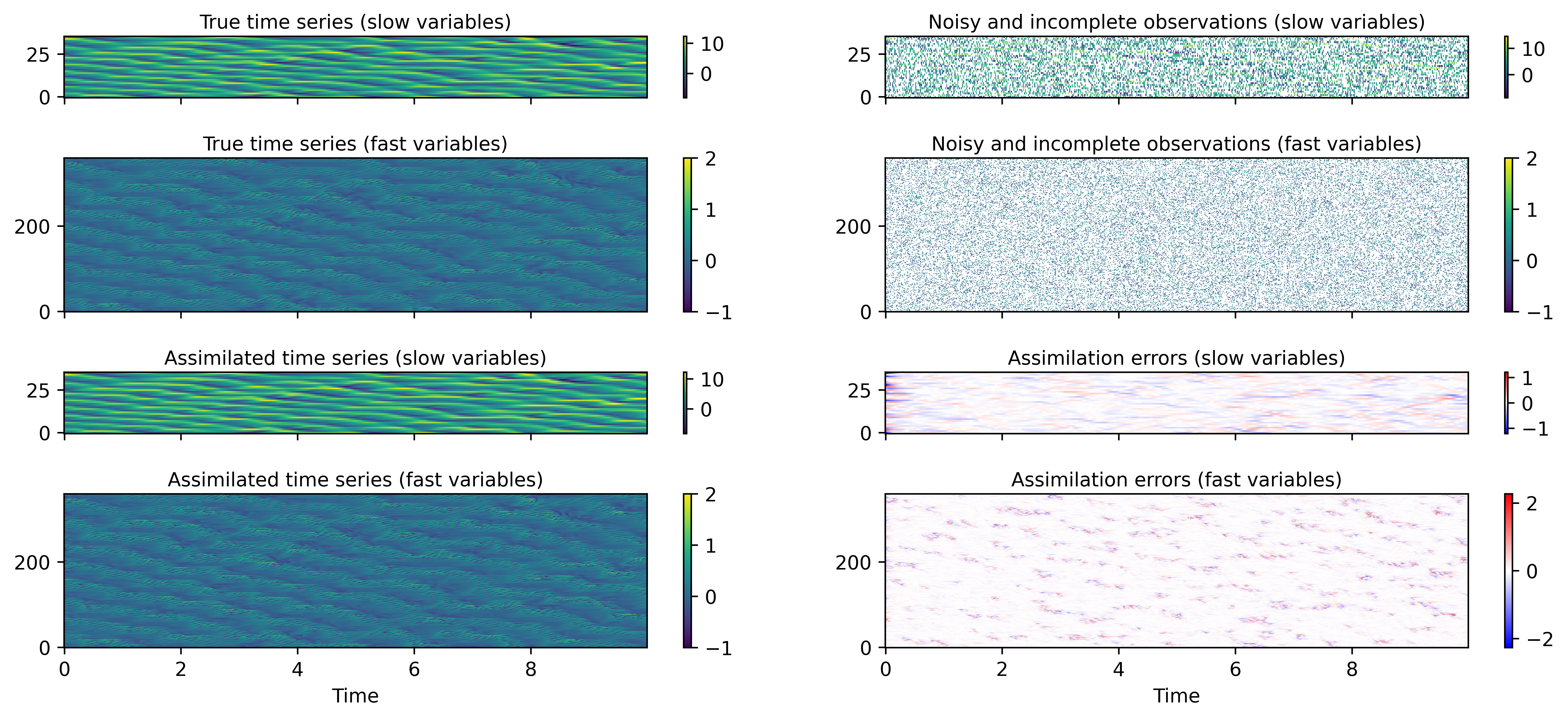}
    \caption{Visualization of the true values, sparse and noisy observations, assimilated time series (with weak-constraint 4D-Var) and assimilation errors for the two-timescale Lorenz-96 system, using observations from both the slow and fast variables.}
    \label{fig:TTS_L96_jointassim}
\end{figure}

Here, we randomly mask~$27$ of the~$36$ slow variables and~$334$ of the~$360$ fast variables at each time step, so that~${75\,\%}$ of the slow and~${90\,\%}$ of the fast variables are unobserved. For the unmasked variables, we add a Gaussian random noise with a standard deviation of~$1$ for the slow variables and~$0.1$ for the fast variables.

We solve this assimilation problem with weak-constraint 4D-Var, initialized with nearest-neighbor interpolation in time. The model error distribution is defined as two separate diagonal Gaussian distributions for the slow and fast variables. Since we know that the fast variables evolve more rapidly than the slow ones and are observed more sparsely, their model error standard deviation is set to 5 times higher than that of the slow variables. This ratio must be selected empirically based on DA performance in cases such as this where model dynamics are precisely known. The results of DA (Fig.~\ref{fig:TTS_L96_jointassim}) show that the reconstructions are qualitatively accurate. Notably, the assimilation errors on the fast variables mostly spike in regions of rapid evolution and remain small elsewhere, while the errors on slow variables always remain relatively small.

\subsubsection{Assimilation of the slow variables, ignoring the effect of the fast ones}
\label{sec:TTL96-2}

We next assimilate the observations of the slow variables under the same conditions as in the previous example, but the fast variables are unobserved and entirely unaccounted for. While groundtruth system states were generated using dynamics $\mathcal M$ from Eq.~\eqref{eq:2TS_L96_slow}-\eqref{eq:2TS_L96_fast}, we assimilate using $\mathcal M$ for the one-scale Lorenz system (Eq.~\eqref{eq:L96}), neglecting the last term of the true state equation~\eqref{eq:2TS_L96_slow}. Since the dynamical model is significantly misspecified in this case, the model error that we use has a much higher standard deviation than the one used in Sect.~\ref{sec:L96_4dvar} and~\ref{sec:TTL96-1}, where the model was actually correct. Specifically, we set it to a Gaussian with mean~$0$ and covariance $10^{-2}\mathbf{I}_n$.

\begin{figure}
    \centering
    \includegraphics[width=1.0\linewidth]{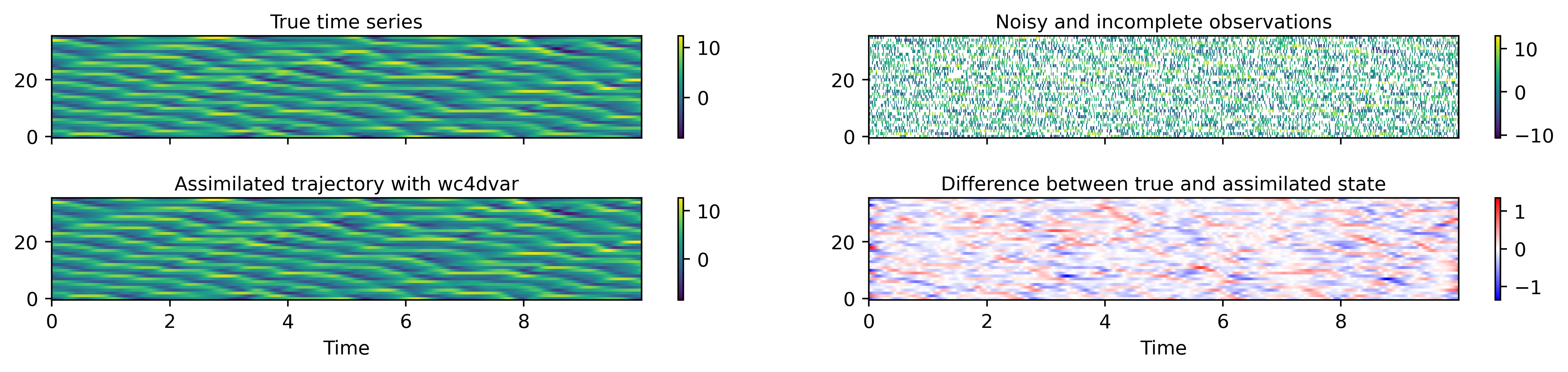}
    \caption{Visualization of the true values, sparse and noisy observations, assimilated time series (with weak-constraint 4D-Var) and assimilation errors for the slow variables of the two-timescale Lorenz-96 system when the influence of the fast variables is completely ignored.}
    \label{fig:TTS_L96_ignoring_fast}
\end{figure}

DA results are shown in Fig.~\ref {fig:TTS_L96_ignoring_fast}. Although the reconstruction remains qualitatively accurate, a comparison with Fig.~\ref{fig:TTS_L96_jointassim} shows that, as expected, ignoring the influence of the fast variables on the slow ones leads to a significant loss of accuracy with respect to the case of a joint assimilation of the two scales.

\subsubsection{Assimilation of the slow variables with joint inference of the unobserved fast ones}
\label{sec:TTL96-3}

Finally, we consider an intermediate case in which the fast variables are completely unobserved, as in the previous experiment, but their influence on the slow ones is still accounted for, and their values are jointly inferred with the slow variables. Thus, this example combines the assimilation method from Sect.~\ref{sec:TTL96-1} with the more challenging observation scenario of Sect.~\ref{sec:TTL96-2}. 
Here, we initialize unobserved fast variables to~$0$ across the observation window, as this value lies within their stationary distribution.

\begin{figure}
    \centering
    \includegraphics[width=1.0\linewidth]{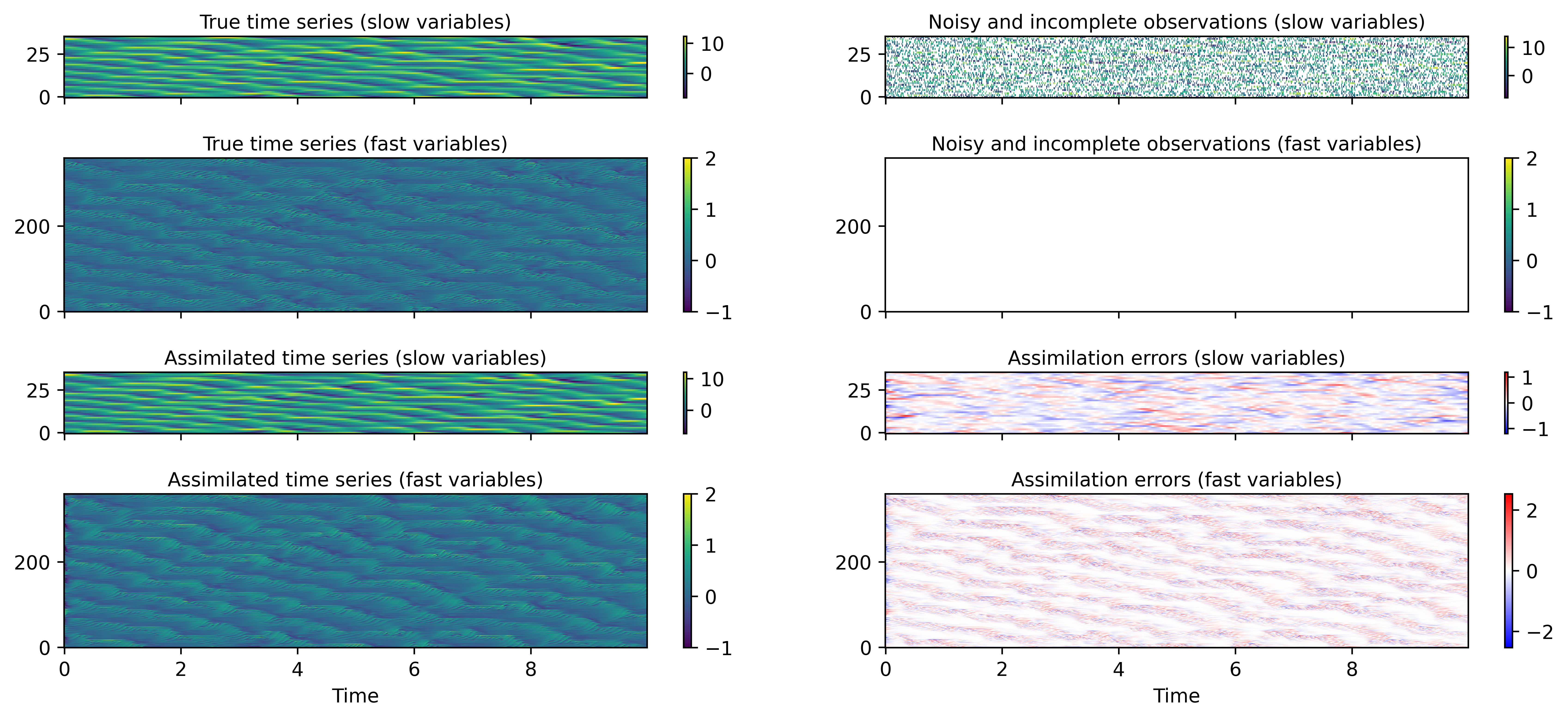}
    \caption{Visualization of the true values, sparse and noisy observations, assimilated time series (with weak-constraint 4D-Var) and assimilation errors for all variables of the two-timescale Lorenz-96 system when the fast variables are not observed but still inferred and taken into account for the assimilation of the system.}
    \label{fig:TTS_L96_inferring_fast}
\end{figure}

The result of the optimization is displayed in Fig.~\ref{fig:TTS_L96_inferring_fast}. The reconstruction is qualitatively accurate for both the observed slow variables and the inferred fast variables. Comparison with Fig.~\ref{fig:TTS_L96_ignoring_fast} reveals that taking the influence of the fast variables on the slow ones into account, even with no additional observations, enables a substantial improvement in the accuracy of the assimilation for the slow variables. Besides, a comparison with Fig.~\ref{fig:TTS_L96_jointassim} shows that, as expected, the absence of observations for the fast variables is still detrimental to the reconstruction quality, for the fast but also for the slow variables.

\subsection{A larger-scale example with multi-layer quasi-geostrophic equations}
\label{sec:QG}

As a larger-scale example, we consider the multi-layer quasi-geostrophic equations, as described by~\cite{hogg_mechanisms_2005}. Since differentiable implementations of these equations are still uncommon, we used (and slightly modified) the implementation proposed by~\cite{thiry_modified_2023}, with equations for pressure and potential vorticity over three layers with background thicknesses 350\,m, 750\,m and 2900\,m. We use an eddy-resolving parameterization on a rectangular domain with a horizontal size of $3.84 \cdot 10^{6}$\,m (width) by $4.8 \cdot 10^{6}$\,m (height). This domain is discretized on a~$769 \times 961$ grid at~$5$\,km resolution. The time step of~$600$ seconds respects the Courant–Friedrichs–Lewy (CFL) condition. The state is defined by the three pressure fields of the vertical layers. Further information on the system and its implementation can be found in~\cite{thiry_modified_2023}.

We work on an observation domain of~$3000$ time steps, corresponding to approximately $21$ days. As in Sect.~\ref{sec:L96}, we generate synthetic observations from the true state trajectory with random masking and Gaussian noise:~${0.5\,\%}$ of the variables are observed, with a noise standard deviation of~$0.05$. To avoid having to split the simulated state trajectory across multiple GPUs (we used one A100 with 80 GB of memory), we use sliding-window strong-constraint 4D-Var with sub-windows of~500 time steps. We initialize each element of the initial state to its first observed value.

For the first sub-window, our background prior distribution is a spherical Gaussian centered on the initial guess and with a standard deviation of~$0.075$ (i.e.~$1.5$ times the standard deviation of the observations). For the following sub-windows, using the \verb|covariance_factor| argument described in Sect.~\ref{sec:sliding_window}, we decrease the background prior s.d. by a factor of 4. For each sub-window, we minimize the 4D-Var cost with 40 Adam optimizer steps using a triangular cyclic learning rate scheduler that varies the learning rate from $2 \times 10^{-3}$ to $5 \times 10^{-3}$ with 20 iterations per cycle. As explained in Sect.~\ref{sec:optimizer}, this gradient descent strategy is readily supported by ADDA, which accepts any combination of PyTorch optimizers and schedulers with prescribed parameters.

\begin{figure}
    \centering
    \includegraphics[width=1.0\linewidth]{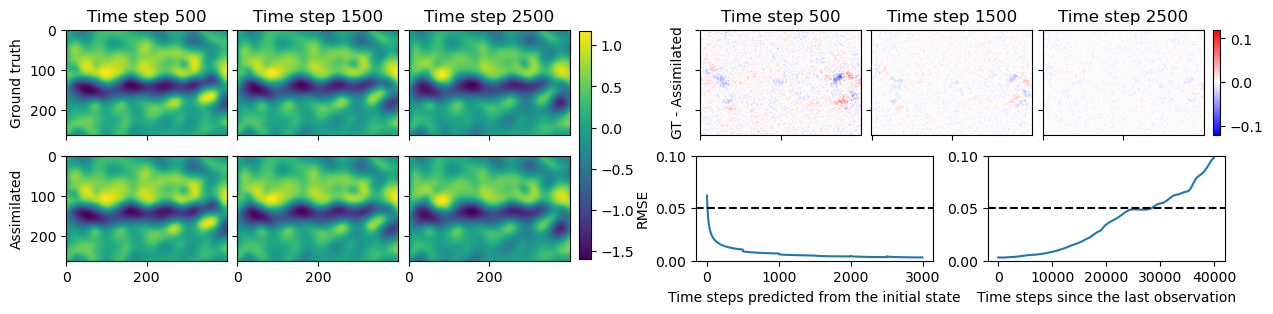}
    \caption{
    Summary of our data assimilation results on the three-layer quasi-geostrophic equations. In the left part of the figure, we show the true and assimilated pressure fields of a central part of the middle layer, for three time steps contained in the observation domain. The corresponding difference fields are displayed in the top-right panel of the figure. On the bottom right part, we plot the global assimilation root mean squared errors as a function of time, both inside of the observation domain (left) and beyond it (right). The dashed horizontal lines correspond to the amplitude of the noise on sparse observations.
    }
    \label{fig:QG_figure}
\end{figure}

Comparing the true and assimilated states at several time points (Fig.~\ref {fig:QG_figure}, left), we observe differences 1-2 orders of magnitude smaller than the states themselves (Fig.~\ref {fig:QG_figure}, top right). RMSE begins slightly above the level of observation noise (dashed black line), and decreases steadily over time (Fig.~\ref {fig:QG_figure}, bottom right). The discontinuities visible in the RMSE every~500 time steps correspond to the transitions between sub-windows of the assimilation.

Finally, to evaluate the robustness of our assimilation, we compare forecasts beyond the final observation to true states. The RMSE gradually increases with the forecast horizon (Fig.~\ref{fig:QG_figure}), but takes more than~$20000$ time steps (significantly longer than the observation domain) to exceed the noise amplitude. An animated version of these DA results is available in the ADDA repository.

\subsection{Vertical diffusion and advection of an inert tracer
\label{sec:4dvargotm}}
To demonstrate how ADDA can incorporate time-varying inputs to the assimilated system, we perform single-window strong- and weak-constraint 4D-Var for the vertical dynamics of an inert tracer in a 1D water column. We consider a horizontally homogeneous water system with time-varying free surface height, such as an estuary subject to tidal cycles.
Since all quantities, including horizontal velocities, temperature and salinity, are assumed to be horizontally homogeneous, we can describe this system in a semi-one-dimensional way where all variables are resolved in only one spatial dimension, but horizontal processes are taken into account when modeling turbulence.
Specifically, we decompose the vertical fluid motion into a mean part which, in our case, is characterized by tidal dynamics, and a fluctuating part that describes turbulent mixing (Reynolds decomposition). This turbulent mixing can effectively be modeled as a form of diffusion.

\begin{figure}[tb]
    \centering
    \includegraphics[width=1.0\linewidth]{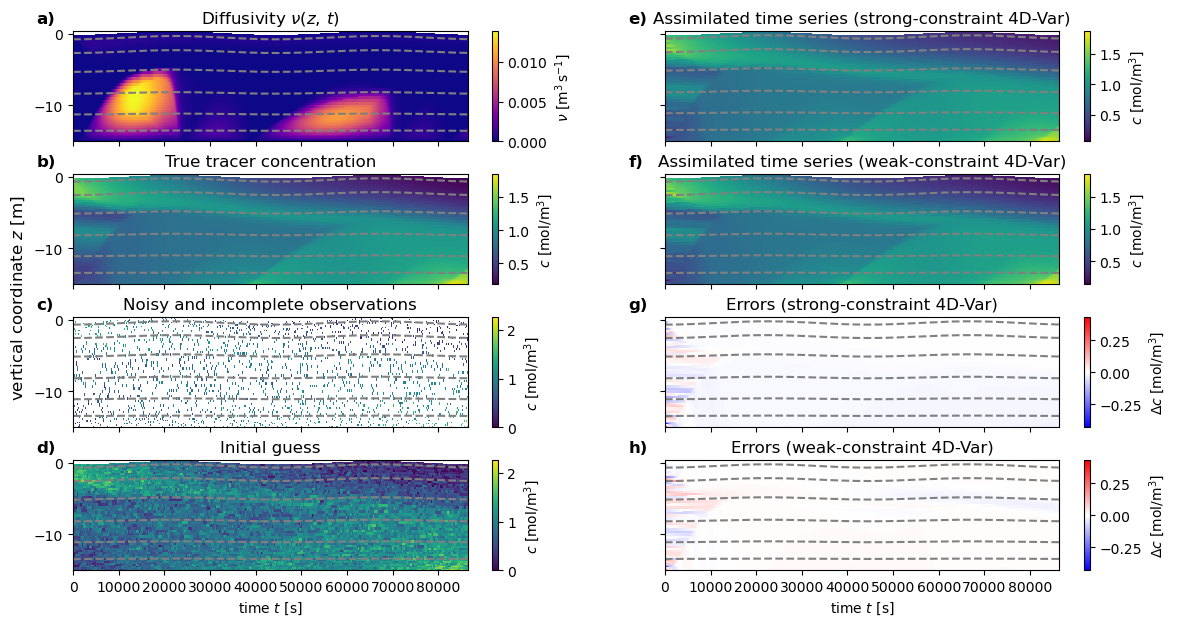}
    \caption{Data assimilation with 4D-Var on an inert tracer in a horizontally homogeneous setting with tides. The tracer concentration~$c$ follows the vertical transport equation~\eqref{eq:advdif}. In all plots, the x-axis is time~$t$ in seconds and the y-axis is the vertical coordinate~$z$ in meters. \textbf{a)} Diffusivity $\nu(z, t)$ as a function of vertical position~$z$ and  time~$t$. \textbf{b)} The tracer concentration $c(z, t)$ simulated by GOTM~\citep{burchard_gotm_1999} acts as a known input to the assimilated system. \textbf{c)} Sparse and noisy pseudo-observations of tracer concentrations with white noise (s.d.~$0.2$,~${10\,\%}$ of values observed). \textbf{d)} Initial guess of the tracer concentrations over space and time using the next observed value. \textbf{e,~f)} Assimilated trajectories using strong-~(e) and weak-constraint~(f) 4D-Var. Strong-constraint 4D-Var was performed using L-BFGS with a learning rate of~0.5 for~5 steps. Weak-constraint 4D-Var used a diagonal model error covariance with diagonal coefficients~${10^{-6}}$, and leveraged L-BFGS with a learning rate of~1.0, for~100 steps. \textbf{g,~h)} Errors of the assimilated trajectories for weak-~(g) and strong-constraint~(h) 4D-Var w.r.t. the true simulated time series.}
    \label{fig:4dvar_inert_tracer}
\end{figure}

Now, we want to perform data assimilation on the time-dependent concentrations of a tracer suspended in the water. The tracer does not interact with other tracers, is neither created nor destroyed, and does not influence the dynamics of the water, but has a constant sinking velocity and otherwise follows the motion of the water. We use the transport equation
\begin{equation}
\label{eq:advdif}
    \dot{c} \; \; = \underbrace{-w_{s} \frac{\partial c}{\partial z}}_{\text{advection (sinking)}} + \underbrace{\frac{\partial}{\partial z}\left(\nu\frac{\partial c}{\partial z}\right)}_{\text{diffusion (turbulence)}}
\end{equation}
with tracer concentration~${c(z, t)}$, constant sinking velocity~$w_{s}$ and turbulent diffusivity~${\nu(z, t)}$. Note that the vertical fluid velocity due to tidal elevation is neglected; instead, the grid moves with the tidal elevation. There is a one-way dependence of the tracer dynamics on the turbulent mixing, which enables us to pre-compute $\nu$ and then use it as a forcing input to our dynamical model for $c$, greatly reducing computational expenses during data assimilation. We perform the turbulence computation with the General Ocean Turbulence Model~(GOTM) introduced by~\cite{burchard_gotm_1999}, specifically with a modified version of the \verb|estuary| test case provided with the model. For further information on this type of scenario, we refer to \citep{burchard_quantifying_2010}. For our actual dynamical model described by Eq.~\eqref{eq:advdif}, we use custom differentiable operators based on the transport routines of GOTM, re-implemented in PyTorch.
Our DA task consists of inferring tracer concentration over space and time, from sparse and noisy observations of the tracer, with both strong- and weak-constraint 4D-Var.


Figure~\ref{fig:4dvar_inert_tracer} shows the application of strong- and weak-constraint 4D-Var to DA for tracer transport. Computation took 101~s for strong- and 158~s for weak-constraint 4D-Var on a 48-core Intel(R) Xeon(R) Platinum 8160 CPU @ 2.10GHz. Both methods reconstruct the tracer concentration time series $c(z, t)$ with high precision. Both strong- and weak-constraint 4D-Var have relatively high reconstruction error at the beginning of the time series, but not at the end. This is likely caused by the stability of the physical system: inaccuracies in the form of small-scale fluctuations in the initial state will disperse and thus become weaker due to diffusion,
limiting the ability of sparse observations to constrain initial states.
This inherent difficulty of initial state estimation of diffusive systems could be addressed by imposing a prior distribution on the initial state. On the other hand, as the initial inaccuracies disappear after a short interval, the objective of state estimation is fulfilled sufficiently well.

\subsection{Assimilation with a neural emulator of Kolmogorov flow 
\label{sec:kfemulator4dvar}}
Neural emulators offer a powerful alternative to classical numerical solvers in modeling physical systems and their improved cost/accuracy tradeoff is demonstrated on large-scale geophysical applications~\citep{kochkov_neural_2023, lam_learning_2023,watt-meyer_ace2_2024, dheeshjith_samudra_2025}. Unlike classical solvers, neural emulators are implemented in automatic differentiation frameworks that natively provide input/output gradients. This makes them more favorable for variational data assimilation than existing complex solvers that lack autodiff capabilities, which would require manual implementation of adjoint routines. By training on system states generated by existing solvers, neural emulators circumvent this need entirely~\citep{nonnenmacher_deep_2021, hatfield_building_2021, frezat_gradient-free_2024}, and integrate naturally into ADDA's framework, which can assimilate any system for which differentiable time stepping is available in PyTorch. To demonstrate this, we train a Fourier neural operator~\citep{li_fourier_2021} following the training procedure described in~\cite{bekar_hybrid_2025} to emulate Kolmogorov flow~\citep{boffetta_two-dimensional_2012}, and use it as the forward model in strong-constraint 4D-Var data assimilation. Kolmogorov flow is governed by the 2D incompressible Navier-Stokes equations with periodic boundary conditions and a forcing consisting of a sinusoidal and a damping term,
\begin{align}
    &\partial_t \mathbf{u} + \nabla \cdot (\mathbf{u} \otimes \mathbf{u}) - \frac{1}{\mathrm{Re}}\Delta \mathbf{u} + \frac{1}{\rho} \nabla p = \mathbf{f} \quad \quad \text{in } \Omega\\
    &\nabla \cdot \mathbf{u} = 0 \quad \quad \text{in } \Omega\\
    &\mathbf{f}=\sin(4y)\hat{\mathbf{e}}_1 - 0.1\mathbf{u}
\end{align}
where $\Omega=[0,2\pi]^2$, $\mathbf{u}=\{u, v\}$ is the velocity field, $\text{Re}$ is the Reynolds number, $\rho$ is the density, $p$ is the pressure field and $\mathbf{f}$ is the forcing. We set $(\rho, \text{Re}, \nu) = (1, 10^3, 10^{-3})$. This system generates statistically stationary flow fields. The direct numerical solver resolves the smallest scale necessary ($\Delta t = 2.1\times10^{-4}$, $\Delta x =3\times10^{-3} $) to simulate the turbulent physics with high fidelity~\citep{kochkov_machine_2021} while the emulator is trained only on the spatiotemporally coarsened system states~$\mathbf{u}$. The training data are temporally coarsened by a factor of approximately~$600$ and spatially coarsened by a factor of~$64$ along both spatial axes, and do not include forcing inputs.

\begin{figure}
    \centering
    \includegraphics[width=\linewidth]{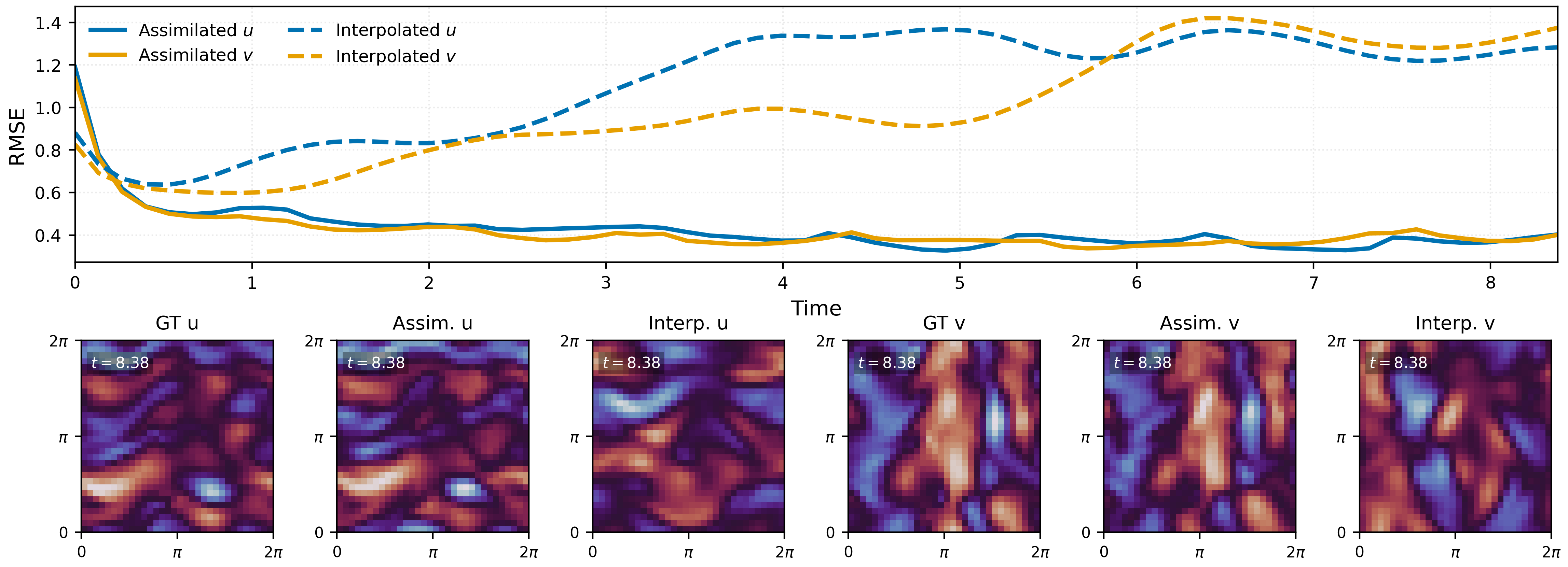}
    \caption{Results of sliding-window strong-constraint 4D-Var on the Kolmogorov flow emulator task. Assimilated states display a lower RMSE and stay closer to the groundtruth trajectory throughout the simulation. Visual inspection of the predicted fields at the end of the assimilation window also demonstrates the improvement.}
    \label{fig:emulator-4dvar-results}
\end{figure}

For the assimilation task, we observe~$5{\,\%}$ of state variables for~64 time steps, with Gaussian observation noise ($\sigma=1$). We apply strong-constraint 4D-Var with non-overlapping~8-step sub-windows (using ADDA's named input \verb|window_shift|). We initialize each element of the system state to its first observed value, and use a spherical Gaussian background prior ($\sigma=2$). We use the same optimizer and scheduler settings as in Sect.~\ref{sec:QG} but, for each sub-window, perform~100 optimization steps with~10 iterations per cycle. Figure~\ref{fig:emulator-4dvar-results} compares the assimilated trajectory to the initialization. DA produces system states that closely follow the groundtruth velocity field, while the initialization exhibits steadily increasing errors and is no longer correlated with the groundtruth fields at the end of the observation domain. This example demonstrates the ease of incorporating neural emulators into ADDA methods, as well as the effectiveness of neural emulators in enabling variational DA for simulations whose gradients are unavailable.

\subsection{Latent data assimilation on the Kuramoto-Sivashinsky system 
\label{sec:latent4dvarks}}
In high-dimensional geophysical problems, such as numerical weather prediction, construction of the background error covariance matrix may be computationally prohibitive~\citep{xiao_vae-var_2024, fan_physically_2026}. While common approaches like localization might mitigate this, likelihood calculations remain costly (see e.g.~\cite{carrassi_data_2018} appendix A). Latent data assimilation (LDA) is a promising approach to addressing this challenge, in which a latent representation of the state space is learned and DA is carried out to identify the latent representations given observations. ADDA's routines for strong-constraint 4D-Var can be used for LDA simply by providing an optional named input \verb|c_proj| of type \verb|callable|. This input receives the decoder of a trained (variational) autoencoder, which maps latent states~$\mathbf{z}$ onto physical states~$\mathbf{x}_t$.

To demonstrate ADDA's straightforward incorporation of LDA, we implement the approach from~\citep{fan_physically_2026}, using the 1D Kuramoto-Sivashinsky (KS) dynamical system~\citep{weinan_gibbsian_2002} as a test case. This system evolves with the following PDE, boundary and initial conditions, resulting in chaotic dynamics:
\begin{equation}
\label{eq:kuramoto}
        \frac{\partial u}{\partial t} = -\frac{\partial^2 u}{\partial x^2} - \frac{\partial^4 u}{\partial x^4} - \frac{1}{2}\left(\frac{\partial u}{\partial x}\right)^2, \quad \quad u(0, t)=u(L,t), \quad \quad     u(x, 0) = \sin(16\pi (x-\theta) / L), \quad \quad \theta\sim U[0,L].
\end{equation}
Initial conditions are sine waves with wavenumber 8 and uniform random phases. We solve the KS equation using a pseudo-spectral solver with an ETDRK2~\citep{cox_exponential_2002} time integrator implemented in PyTorch following the approach of \cite{koehler_apebench_2024}. The domain length is set to $L=100$ with spatial resolution $\Delta x \approx 0.78$ and time step $\Delta t=0.1$. As we aim to carry out DA for chaotic trajectories from the stationary distribution on the KS system's attractor, we generate groundtruth trajectories by running the solver for~$1600$ steps and discarding the first~$800$. We implement the strong-constraint 4D-Var LDA formulation from \cite{fan_physically_2026},
\begin{equation}
\label{eq:latentdaloss}
    \underset{\mathbf{z} \in \mathbb{R}^{s}}{\text{min}} J(\mathbf{z}) = \frac{1}{2}||\mathbf{z} - \mathbf{z}_B||^2_{\boldsymbol{\Sigma}_{B_{\mathbf{z}}}^{-1}} + \frac{1}{2} \sum_{t=0}^T||\mathbf{y}_t - \mathcal{H}_t(\mathcal{M}^{(t)}(\mathcal{D}(\mathbf{z})))||^2_{\boldsymbol{\Sigma}_{\boldsymbol\eta_t}^{-1}},
\end{equation}
where~$\mathbf{z}$ is the initial latent state, $\boldsymbol{\Sigma}_{B_{\mathbf{z}}}$ is the background error covariance matrix in the latent space and~$\mathcal{D}$ is the decoder mapping latent states to physical states. As argued in~\cite{fan_physically_2026}, constructing the background error covariance matrix in latent space is more tractable than in data space, and they empirically show that it becomes approximately diagonal. They specify this matrix manually in the latent space of a regular auto-encoder, but we take a different approach and train a $\beta$-VAE instead~\citep{xiao_vae-var_2024}. Since the VAE objective explicitly regularizes the latent variable toward a normal distribution~\citep{higgins_-vae_2017} and our training data are drawn from the stationary regime of the KS system, we define the background prior on the latent states also as a normal distribution, i.e., $\boldsymbol{\Sigma}_{B_{\mathbf{z}}}$ an identity matrix and~$\mathbf{z}_B$ a vector of zeros. 

\begin{figure}
    \centering
    \includegraphics[width=1.0\linewidth]{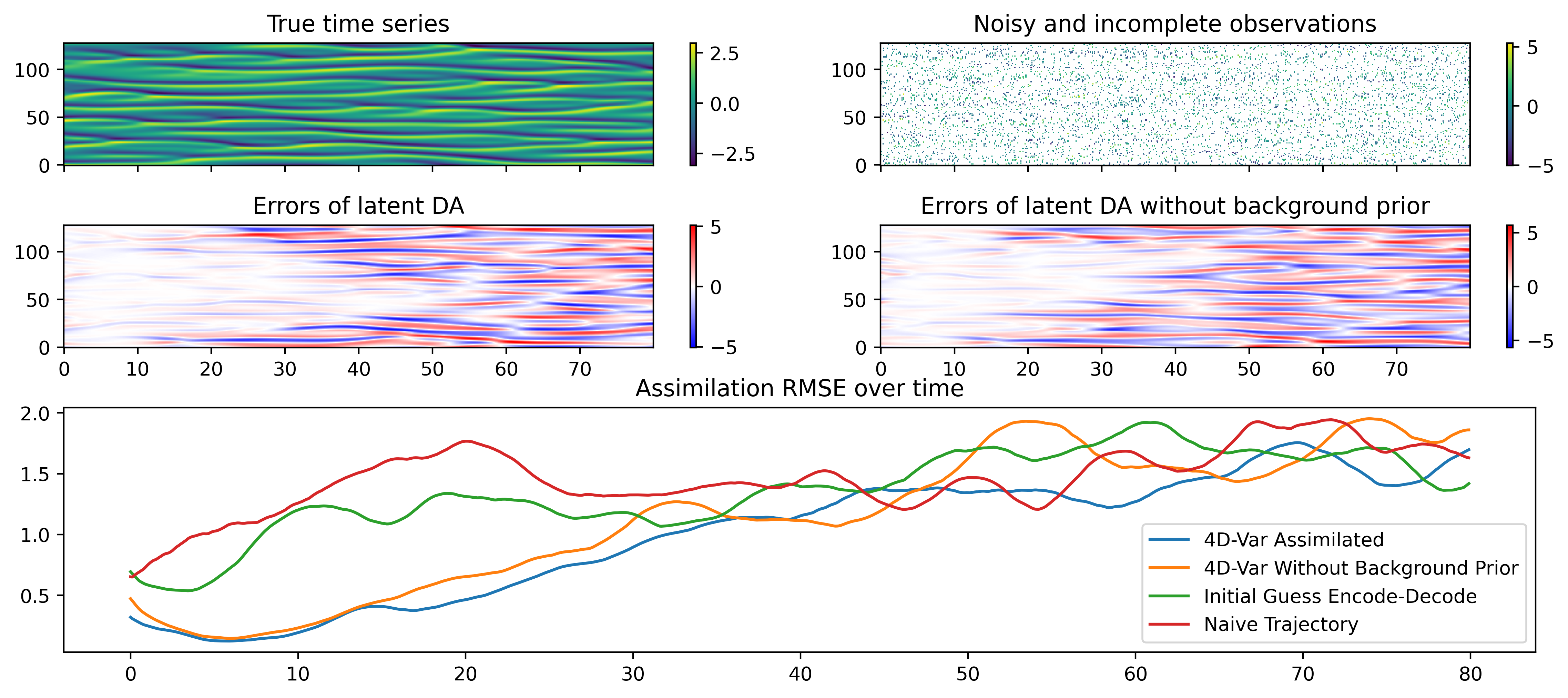}
    \caption{Assimilated results with strong-constraint 4D-Var with latent data assimilation on the Kuramoto-Sivashinsky system. Using a background prior in the latent space results in more accurate assimilation.}
    \label{fig:lda-results}
\end{figure}

We assimilate over a window of~100 time steps, i.e. 10 time units, observing~${5\,\%}$ of the state variables, with Gaussian white noise of s.d. $1.0$. The latent state~$\mathbf{z}$ is initialized by encoding a state $\mathbf{x}_0$ obtained by nearest-neighbor interpolation over time. Optimization is performed using L-BFGS for~20 steps with a learning rate of~0.1. 

We compare four configurations: LDA with background prior, LDA without background prior, rolling out the encode-decode initialization of $\mathbf{x}_0$ without assimilation, and rolling out the naive initialization without assimilation. Note that even without an explicit prior term, the trained decoder provides an implicit (or "soft") prior: because it has learned to represent the training data distribution, optimizing through the decoder naturally biases the solution toward the learned data manifold rather than arbitrary states. As shown in Fig.~\ref{fig:lda-results}, LDA with background prior performs the best, while the second best model is the LDA without background prior, followed by the encode-decode baseline. We believe that encoding and decoding act as a projection of the noisy input onto the manifold of the stationary dynamics. The nearest-neighbor-in-time initialization performs the worst. All the trajectories diverge after~$400$ steps due to the chaotic nature of the KS system.

\subsection{Strong-constraint 4D-Var on the 2D Kuramoto-Sivashinsky system with a JAX-Torch bridge
\label{sec:ksjax4dvar}}
JAX \citep{bradbury_jax_2018} has a growing ecosystem of differentiable scientific software \citep{kochkov_machine_2021, bezgin_jax-fluids_2023, holl_phiflow_2024, koldunov_fesom2-jax_2026}, while PyTorch offers a more established deep learning ecosystem. We implement ADDA in PyTorch to leverage this maturity and its wider support for deep learning-based data assimilation methods. Since both frameworks support automatic differentiation, we bridge a JAX simulation code with ADDA for variational data assimilation. 

\begin{figure}
    \centering
    \includegraphics[width=\linewidth]{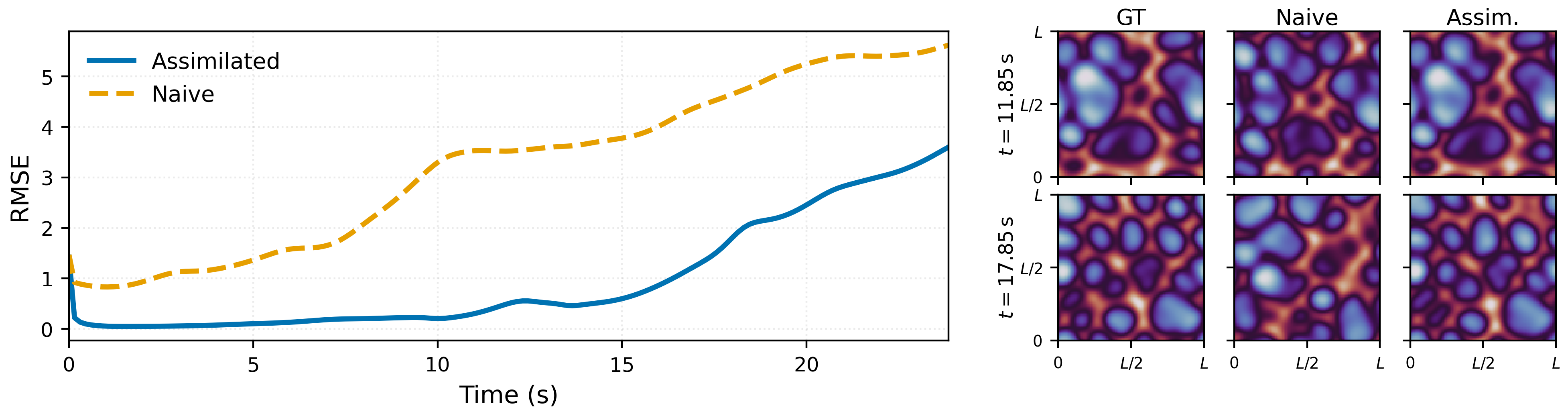}
    \caption{Assimilated results with strong-constraint 4D-Var on the 2D KS system. The RMSE of the predictions over time (left) and snapshots of the field variable at $t=11.85s$ and $t=17.85s$ (right). The predictions from the assimilated initial condition remain lower throughout the prediction window.}
    \label{fig:ks2d-4dvar-jax}
\end{figure}

The bridge is implemented as a custom \verb|torch.autograd.Function|, where \verb|jax.vjp| passes gradients from JAX to PyTorch, while dlpack performs zero-copy conversion of arrays between the two frameworks. This introduces computational overhead relative to a native PyTorch implementation, but avoids redundant re-implementation of scientific software that already supports automatic differentiation. Coupling frameworks in this way might be a practical alternative to re-implementation; it extends ADDA to the growing body of JAX-based differentiable solvers and increases its practicality as research software.
To demonstrate this, we use Exponax \citep{koehler_apebench_2024}, a differentiable spectral solver library built in JAX, with the 2D-KS equation as a benchmark:
\begin{equation}
\label{eq:kuramoto2d}
        \frac{\partial u}{\partial t} + \frac{1}{2} \| \nabla u \|_2^2 + \Delta u + \Delta^2 u = 0
, \quad \quad u(0, y, t)=u(L_x, y, t), \quad \quad u(x, 0, t)=u(x, L_y, t).
\end{equation}
At this domain size, the system exhibits chaotic dynamics. The initial condition is sampled as $u(x,y,0) \sim \mathcal{N}(0,1)$, with domain length $L_x=L_y=30$, spatial resolution $\Delta_x=\Delta_y=0.3$, and time step $\Delta_t=0.15$, integrated using the ETDRK2 solver. The initial 1000 transient time steps are discarded, and the following 160 time steps form the dataset. We observe $25\,\%$ of state variables for 16 time steps with Gaussian observation noise ($\sigma=1$) and apply strong-constraint 4D-Var, initializing each state element to its first observed value. We use the same optimizer and scheduler settings as in Sect.~\ref{sec:latent4dvarks}, optimizing for 5 steps. Figure~\ref{fig:ks2d-4dvar-jax} shows the RMSE of the assimilated versus initialized trajectories across 160 time steps and contours at two time points; the assimilated states exhibit lower prediction error throughout the prediction window and predictions from the assimilated initial condition remain close to the ground truth.

\section{Discussion}\label{sec:discussion}

We consider this work to mark a shift in the relative technical difficulties of implementing variational and ensemble-based data assimilation methods. Historically, 4D-Var has been considered more powerful but also more difficult to develop and maintain than the ensemble Kalman filter and smoother (see e.g.~\cite{altaf_reduced_2013,carrassi_data_2018}) due to the necessity of deriving an adjoint model of the studied dynamical system to compute its gradients. It has long been envisioned that automatic differentiation could be a crucial tool for overcoming this difficulty (e.g.~\cite{daescu_adjoint_2000,castaings_automatic_2006}), yet a flexible and extensive software basis for performing data assimilation with the help of automatic differentiation was still lacking. Another important advantage of automatic differentiation for data assimilation is the use of GPUs for efficient computation, which is straightforward in autodiff frameworks such as PyTorch. We hope that the present work will contribute to a large shift of data assimilation practice towards automatic differentiation. Concretely, the main difficulty that remains unaddressed by our work is that many of the geoscientific dynamical systems on which data assimilation is performed operationally rely on complex implementations in old but computationally efficient programming languages such as Fortran, which do not enable automatic differentiation. This difficulty can be tackled with two different strategies: first, one can directly re-implement the systems in an automatic differentiation framework. This approach, known as differentiable physics, has attracted increasing attention~\citep{shen_differentiable_2023} and has already achieved significant success in a range of applications~\citep{gelbrecht_differentiable_2023,sapienza_differentiable_2025}. A popular alternative is to train an emulator, i.e., a neural network designed to approximate the system dynamics, typically at a significantly lower cost and allowing for differentiation by design~\citep{hatfield_building_2021, maulik_efficient_2022, seabra_ai_2024}.

The potential extensions of our work are numerous. First, it should be noted that the field of data assimilation is extremely extensive and that many popular methods, either classical (such as various ensemble Kalman filter variants and weak-constraint 4D-Var schemes) or machine/deep learning-based (e.g., analog data assimilation~\citep{lguensat_analog_2017}, 4DVarNet~\citep{fablet_learning_2021} and score-based data assimilation~\citep{rozet_score-based_2023}), are not yet implemented in ADDA. Support for these techniques could open the way to establishing an extensive benchmark of data assimilation methods. Besides data assimilation itself, the problem of joint state and parameter estimation (discussed in e.g.~\cite{bocquet_joint_2013, zinchenko_combined_2024}) is a natural and straightforward extension of our work, as automatic differentiation could certainly be beneficial in this type of problem as well.

\section{Conclusion}
\label{sec:conclusion}

We have introduced ADDA, a software package for performing data assimilation in a large variety of contexts. Our numerous examples showcase the versatility of ADDA, which can handle e.g. states with heterogeneous fields, dynamics with forcing inputs and observations irregularly sampled in time. The modular aspect of ADDA further enables easy adaptation to new use cases, including custom dynamics, observation operators and probability distributions. A core feature of ADDA is that it leverages automatic differentiation in PyTorch, while also offering compatibility with an extensive corpus of existing differentiable simulation in JAX. ADDA fulfills a longstanding need to perform variational data assimilation without going through the tedious task of implementing an adjoint model for every dynamical model in addition to the implementation of the model itself. Now, one only needs to provide a PyTorch or JAX implementation of the dynamical model, and the adjoint is then obtained through automatic differentiation. In addition, recent methods involving neural networks, such as latent data assimilation and 4D-Var with a deep emulator, are straightforward with ADDA, as shown in our experiments. We expect this work to enable a more widespread usage of variational data assimilation with differentiable physics simulators as well as deep emulators, and extensive comparisons between different dynamical models and optimization methods.

\bibliographystyle{unsrtnat}
\bibliography{references.bib}

\end{document}